\pdfoutput=1 
\documentclass[10pt,twocolumn,journal]{IEEEtran}
\usepackage{latexsym}
\usepackage{verbatim}
\usepackage{amsfonts}
\usepackage{amsbsy}
\usepackage{amssymb}
\usepackage{flushend,cuted}
\usepackage{amsmath,amssymb}
\usepackage{times}
\usepackage{graphicx}
\usepackage{enumerate}
\usepackage[usenames]{color}
\usepackage[dvips]{pstcol}
\usepackage{epstopdf}
\usepackage{cite}
\usepackage{amsmath}
\usepackage{amssymb}
\usepackage{amsfonts}
\usepackage{graphicx}
\usepackage{epsfig}
\usepackage{psfrag}
\usepackage{xcolor}
\usepackage{amsfonts, bm}
\usepackage{epstopdf}
\usepackage{cite}
\usepackage{color}
\usepackage{xcolor}
\usepackage{subfig}
\usepackage{algorithm}
\usepackage{algorithmicx}
\usepackage{algpseudocode}
\usepackage{amsmath}
\usepackage{multirow}
\input epsf
\begin{document}
	\title{Mixed-Timescale Deep-Unfolding for Joint Channel Estimation and Hybrid Beamforming}


	\author{ Kai Kang, Qiyu Hu, \textit{Student Member, IEEE,} Yunlong Cai, \textit{Senior Member, IEEE,} \\Guanding Yu, \textit{Senior Member, IEEE,} Jakob Hoydis, \textit{Senior Member, IEEE,} and Yonina C. Eldar, \textit{Fellow, IEEE}
		\thanks{
			K. Kang, Q. Hu, Y. Cai, and G. Yu are with the College of Information Science and Electronic Engineering, Zhejiang University, Hangzhou 310027, China (e-mail: kangkai@zju.edu.cn; qiyhu@zju.edu.cn; ylcai@zju.edu.cn; yuguanding@zju.edu.cn).
			J. Hoydis is with NVIDIA, 06906 Sophia Antipolis, France (e-mail:
			jhoydis@nvidia.com).
			Y. C. Eldar is with the Department of Mathematics and Computer Science, Weizmann Institute of Science, Rehovot 7610001, Israel (e-mail: yonina.eldar@weizmann.ac.il).
		}
	}

	\maketitle
	\vspace{-3.7em}
	\begin{abstract}
		
		In massive multiple-input multiple-output (MIMO) systems, hybrid analog-digital beamforming is an essential technique for exploiting the potential array gain without using a dedicated radio frequency chain for each antenna. However, due to the large number of antennas, the conventional channel estimation and hybrid beamforming algorithms generally require high computational complexity and signaling overhead. In this work, we propose an end-to-end deep-unfolding neural network (NN) joint channel estimation and hybrid beamforming (JCEHB) algorithm to maximize the system sum rate in time-division duplex (TDD) massive MIMO. Specifically, the recursive least-squares (RLS) algorithm and stochastic successive convex approximation (SSCA) algorithm are unfolded for channel estimation and hybrid beamforming, respectively. In order to reduce the signaling overhead, we consider a mixed-timescale hybrid beamforming scheme, where the analog beamforming matrices are optimized based on the channel state information (CSI) statistics offline, while the digital beamforming matrices are designed at each time slot based on the estimated low-dimensional equivalent CSI matrices. We jointly train the analog beamformers together with the trainable parameters of the RLS and SSCA induced deep-unfolding NNs based on the CSI statistics offline. During data transmission, we estimate the low-dimensional equivalent CSI by the RLS induced deep-unfolding NN and update the digital beamformers. In addition, we propose a mixed-timescale deep-unfolding NN where the analog beamformers are optimized online, and extend the framework to frequency-division duplex (FDD) systems where channel feedback is considered. Simulation results show that the proposed algorithm can significantly outperform conventional algorithms with reduced computational complexity and signaling overhead.
	\end{abstract}

	\IEEEpeerreviewmaketitle
	
	\section{Introduction}
	\label{sec:intro}
	 Thanks to large-scale spatial multiplexing and highly directional beamforming, massive multiple-input multiple-output (MIMO) has been recognized as a pivotal technology for improving system reliability and data rate \cite{mmWave1, mmWave2, mmWave3, MIMO1, MIMO2}. However, due to the exorbitant cost and energy consumption of radio frequency (RF) chains and analog-to-digital converters, the employment of conventional fully-digital beamforming is impractical with current technologies. Thus, hybrid analog-digital beamforming which requires a smaller number of RF chains has received great attention\cite{HB1, HB2}. There have been a number of algorithms proposed for hybrid beamforming and channel estimation in massive MIMO systems \cite{Hybrid0, Hybrid1, Hybrid3, Hybrid4, Hybrid5, Hybrid6, Hybrid7,CE2, CE3, CE4, CE5, CE6}. These approaches typically require high complexity and signaling overhead.  Moreover, these two modules are generally designed separately, which may result in performance loss.  
	
	 We consider a joint design of channel estimation and hybrid beamforming with low-complexity and reduced overhead. 
	 A number of previous algorithms have been proposed for hybrid beamforming in \cite{Hybrid0, Hybrid1, Hybrid3, Hybrid4, Hybrid5, Hybrid6, Hybrid7}. In \cite{Hybrid1}, the authors proved that if the RF chain equipped with the hybrid beamforming structure is twice the total number of data streams, the performance approaches that of fully-digital beamforming. In \cite{Hybrid3} and \cite{Hybrid4}, a hybrid beamforming framework was suggested for improving the bit error rate and system sum rate performance, respectively. Considering hardware constraints, codebook-based methods for hybrid beamforming were investigated in \cite{Hybrid5, Hybrid6, Hybrid7}. In particular, a hierarchical codebook design for hybrid beamforming was proposed by \cite{Hybrid5} while a codebook-based RF precoding designed to maximize the spectral efficiency and energy efficiency simultaneously was designed in \cite{Hybrid6}. Channel estimation plays an important role in hybrid beamforming design\cite{CE1, CE2, CE3, CE4, CE5, CE6}. The authors of \cite{CE2} developed an effective algorithm that uses an hidden Markov model (HMM) for sparse channel estimation. In \cite{CE3}, the authors proposed a compressive sensing method for channel estimation by exploiting the spatial sparsity. A recursive least-squares (RLS) adaptive estimation algorithm was developed for MIMO interference channels in \cite{CE6}, which provides low computational complexity and can track the time-varying channels as the environment changes. 
	
 	Conventional single-timescale hybrid beamformers are optimized based on the high-dimensional full channel state information (CSI), which leads to large signaling overhead and transmission delay. To address these issues, several hybrid beamforming algorithms under the mixed-timescale scheme have been investigated in \cite{Twotime1, Twotime2, Twotime3}. In this approach, long-term analog beamformers are optimized based on the channel statistics while the short-term digital beamformers are updated based on the reduced-dimensional CSI. However, these algorithms are challenging to implement in practice owing to the large number of iterations for convergence and high computational complexity operations, such as matrix inversion in each iteration.
 	
	In recent years, deep learning techniques have been widely applied in wireless communications such as channel estimation \cite{DL1}, signal detection \cite{DL11, DL2, DL22}, and CSI feedback in MIMO systems \cite{DL3}. Compared to traditional algorithms, deep learning-based techniques have much lower computational complexity and often do not require CSI. In \cite{DL_BF1, DL_BF2, DL_BF3, E2E}, the authors designed hybrid beamforming by employing convolutional neural networks (CNNs) and multi-layer perception (MLP) which are referred to as black-box neural networks (NNs). However, these NNs have poor interpretability and many samples are required for training. Deep-unfolding NNs have been recently receiving growing interest in various areas \cite{DU}. This approach unfolds iterative algorithms into layer-wise networks and introduces trainable parameters to improve system performance. Compared with black-box NNs, deep-unfolding NNs are more interpretable and require less training data, and have much lower computational complexity compared to traditional algorithms with comparable performance. Deep-unfolding NNs have been applied in communications \cite{DU9}, for example, resource allocation \cite{DU00, DU0}, detection \cite{DU11, DU1,DU2, DU3}, channel estimation \cite{DU4, DU5}, and transceiver design \cite{DU6, DU7, DU8, DU10}. In \cite{DU2}, the authors proposed a symbol detector named ViterbiNet, which integrates black-box NNs into the Viterbi algorithm.
	
    In prior works, deep-unfolding NNs are employed for a single module design.
    In this work, we propose an end-to-end deep-unfolding framework for joint channel estimation and hybrid beamforming (JCEHB) design in time-division duplex (TDD) massive MIMO systems to maximize the system sum rate. To reduce the signaling overhead, we employ the mixed-timescale hybrid beamforming scheme where analog beamformers are optimized offline. In addition, we extend the framework to other application scenarios.
	
	The main contributions of this work are as follows.
	\begin{itemize}
	\item We propose an end-to-end mixed-timescale deep-unfolding framework for maximizing the system sum rate in massive MIMO, which jointly designs channel estimation and hybrid beamforming. 
	\item We develop a RLS algorithm induced channel estimation deep-unfolding NN (CEDUN) and an SSCA algorithm induced hybrid beamforming deep-unfolding NN (HBDUN). For the CEDUN, we design the pilot training module and unfold the RLS algorithm into a layer-wise NN with introduced trainable parameters. For the HBDUN, we propose a stochastic successive convex approximation (SSCA) algorithm induced deep-unfolding NN, where the high computational complexity operations are replaced by trainable parameters.
	\item Under the mixed-timescale scheme, we develop a two-stage joint training method for the deep-unfolding NNs, where the analog beamformers are treated as trainable parameters and optimized offline based on the channel statistics. During the data transmission stage, we fix the analog beamformers and only estimate the low-dimensional equivalent CSI by the CEDUN to update the digital beamformers. When channel statistics change, we employ transfer learning to fine-tune the deep-unfolding NN. 
	\item We extend our framework for the following different application scenarios: (i) An online mixed-timescale scheme where the analog beamformers are optimized in an online manner; (ii) Frequency-division duplex (FDD) system that incorporates channel quantization and feedback. We propose an end-to-end deep-learning based framework where deep-unfolding NNs are designed for channel estimation and hybrid beamforming and black-box NNs are designed for channel quantization and feedback, respectively. 
	\item We provide detailed analysis of the performance and computational complexity of the proposed deep-unfolding algorithm. Simulation results show that our proposed deep-unfolding  can significantly outperform conventional RLS and SSCA algorithms with reduced complexity.
	\end{itemize}

	The rest of the paper is structured as follows. Section \ref{Section2:system}
	introduces the system model and problem formulation. Section \ref{framework} proposes the mixed-timescale deep-unfolding framework and presents the joint training method. Section \ref{Section3:CE} develops the RLS induced deep-unfolding NN for channel estimation and Section \ref{HB} proposes the SSCA deep-unfolding NN for hybrid beamforming design. Section \ref{extension} extends the deep-unfolding framework for different application scenarios. Section \ref{analysis} analyzes the computational complexity and performance of the proposed algorithm. Section \ref{Section6:simulations} presents simulation results and conclusions are drawn in Section \ref{Section7:conclusion}.

	Throughout the paper we use the following notations. Scalars, vectors and matrices are respectively denoted by lower case, boldface lower case and boldface upper case letters;
	$\mathbf{I}$ represents an identity matrix and $\mathbf{0}$
	denotes an all-zero  matrix.
	For a  matrix $\mathbf{A}$, ${{\bf{A}}^T}$, $\mathbf{A}^*$, ${{\bf{A}}^H}$, and $\|\mathbf{A}\|$ denote its transpose, conjugate, conjugate transpose and Frobenius norm, respectively. For a square matrix $\bf{A}$, $\textrm{Tr} \{\bf{A}\}$ is its trace.
	For a vector $\mathbf{a}$, $\|\mathbf{a}\|$ represents its Euclidean norm,
	$\mathbb{E}\{\cdot\}$ denotes the statistical expectation, and
	$\Re\{\cdot\}$ ($\Im\{\cdot\}$) are
	the real (imaginary) part of a variable.
	The operator $\textrm{vec}( \cdot )$ stacks the elements of a matrix in a column vector,
	$|  \cdot  |$ denotes  the absolute value of a complex scalar, and
	${\mathbb{C}^{m \times n}}\;({\mathbb{R}^{m \times n}})$ are the space of ${m \times n}$ complex (real) matrices.
	
	\vspace{0.6em}
	\section{ System Model and Problem Formulation}
	\label{Section2:system}
	In this section, we first introduce the system model for downlink massive MIMO and then formulate our problem mathematically.
	\subsection{System Model}
	\subsubsection{Signal Model} 
\begin{figure}[t]
		\begin{centering}
			\includegraphics[width=0.5\textwidth]{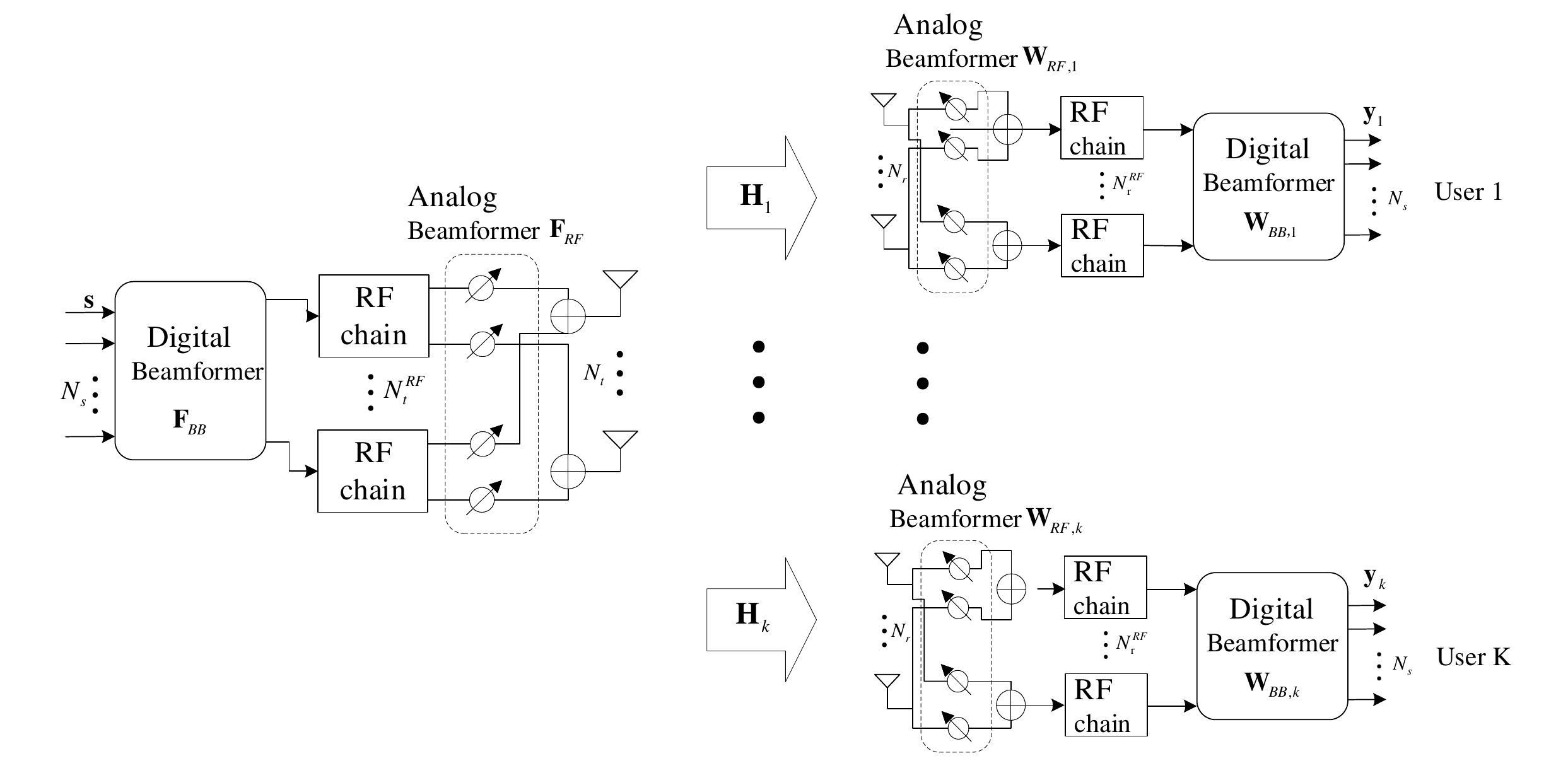}
			\par\end{centering}
		\caption{Downlink massive MIMO system with hybrid beamforming.}
		\label{HybridModel}
	\end{figure}

	Consider a downlink massive MIMO system working in TDD mode as shown in Fig. \ref{HybridModel}. The base station (BS) is equipped with $N_{t}$ transmit antennas and $N^{RF}_{t}$ RF chains, sending $N_{s}$ data streams to each user at the receiver with $ K $ users, where $KN_{s}\leq N^{RF}_{t} \leq N_{t}$. Each user is equipped with $N_r$ receive antennas and $N^{RF}_{r}$ RF chains, where $N_{s}\leq N^{RF}_{r} \leq N_{r}$. At the transmitter, the RF chains are connected with a network of phase shifters that expands the $N^{RF}_{t}$ digital outputs to $N_{t}$ precoded analog signals feeding the transmit antennas. Similarly, at the receiver, the $N_{r}$ receive antennas are followed by a network of phase shifters that feeds the $N^{RF}_{r}$ RF chains.
	The BS sends $ N_{s} $ data streams to user $ k \in \mathcal{K} \triangleq \{1, \ldots, K\} $, denoted as $ \mathbf{s}_{k} \in\mathbb{C}^{N_{s} \times 1} $. Through the beamformers at the BS, the signal $ \mathbf{u}_k \in \mathbb{C}^{N_t \times 1} $ for user $ k $ can be written as
	\begin{equation}
		\mathbf{u}_{k} = \sqrt{P}\mathbf{F}_{RF,k}\mathbf{F}_{BB,k}\mathbf{s}_{k}, \label{u_{k}}
	\end{equation}
	where $ P $ denotes the transmit power of the BS, $\mathbf{F}_{RF,k}\in\mathbb{C}^{N_{t}\times N_{t}^{RF}}$ is the analog beamformer which is subject to a unit modulus constraint, i.e., $ |[\mathbf{F}_{RF}]_{i,j}| = \frac{1}{\sqrt{N_{t}}},\forall i,j $, and $\mathbf{F}_{BB,k} \triangleq [\mathbf{f}_{BB,k,1}, \ldots, \mathbf{f}_{BB,k,Ns}]  \in\mathbb{C}^{N_{t}^{RF}\times N_{s}}$ is the digital beamformer. The digital precoder $ \mathbf{F}_{BB,k} $ is normalized as $ \|\mathbf{F}_{RF,k}\mathbf{F}_{BB,k}\|_{F}^{2} = N_{s} $ to ensure that the power constraint is satisfied at the BS.
	
    After passing through the channel and the beamformers at the user, the received signal vector for user $ k $ is
	\begin{equation}
	\begin{aligned}
	&\mathbf{y}_{k} = \sqrt{P}\mathbf{W}_{BB,k}^{H}\mathbf{W}_{RF,k}^{H}\mathbf{H}_{k}\mathbf{F}_{RF,k}\mathbf{F}_{BB,k}\mathbf{s}_{k} 
	+ \sqrt{P}\mathbf{W}_{BB,k}^{H}\\&\mathbf{W}_{RF,k}^{H}\mathbf{H}_{k}\!\!\!\!\!\sum\limits_{m=1,m \neq k}\limits^{K}\!\!\!\!\mathbf{F}_{RF,m}\mathbf{F}_{BB,m}\mathbf{s}_{m}\!
	+ \!\mathbf{W}_{BB,k}^{H}\mathbf{W}_{RF,k}^{H}\mathbf{z}_{k}, \label{y_k}
	\end{aligned}
	\end{equation}
	where ${\mathbf{H}_k \in \mathbb{C}^{N_r \times N_t}}$ represents the channel matrix, ${\bm{W}_{BB,k} \triangleq [\mathbf{w}_{BB,k,1}, \ldots, \mathbf{w}_{BB,k,Ns}] \in {\mathbb{C}^{N_r^{RF} \times N_s}}}$  and 
	${\bm{W}_{RF,k} \in {\mathbb{C}^{N_r \times N_r^{RF}}}}$ denote the digital and analog beamformers for user $ k $, respectively, and ${\mathbf{z}_k \in \mathbb{C}^{N_r \times 1}} \sim {\mathcal{CN}}$ $({\mathbf{0}}, \sigma^2_k\mathbf{I})$ is the additive white Gaussian noise (AWGN) with $\sigma_k$ denoting the noise power. Similar to $ \mathbf{F}_{RF} $, the analog combiner satisfies a unit modulus constraint $ |[\mathbf{W}_{RF}]_{i,j}| = \frac{1}{\sqrt{N_{r}}},\forall i,j $.
	Using (\ref{y_k}), given perfect knowledge of the equivalent channel, the signal-to-interference-plus-noise ratio (SINR) for stream $ l $ of user $ k $ is given as
	\begin{equation}
	\Gamma_{k, l}\! =\! \frac{|\mathbf{w}^H_{BB,k,l}\!{\mathbf{H}}_{eq,k}\!\mathbf{f}_{BB,k,l}|^2}
	{\mathop{\!\!\sum \limits_{i=1}^K \!\sum \limits_{j=1}^{N_{s}}} \limits_{(i,j \neq k,l)}\!\!\! |\mathbf{w}^H_{BB,k,l} \!{\mathbf{H}}_{eq,k}\! \mathbf{f}_{BB,i,j}|^2 
		\!\!+ \!\!\frac{\sigma^2_k}{P} \|\!\mathbf{w}^H_{BB,k,l}\! \mathbf{W}^H_{RF,k}\|^2}, \label{SINR}
	\end{equation}
    where ${\mathbf{H}}_{eq,k} = {\mathbf{W}}_{RF,k}^H \mathbf{H}_{k} {\mathbf{F}}_{RF,k}\in\mathbb{C}^{N_{r}^{RF} \times N_{t}^{RF}}$ denotes the low-dimensional equivalent CSI matrix. {The system sum rate is $ \sum \limits_{k=1}^K \sum \limits_{l=1}^{N_{s}} \log(1 + \Gamma_{k, l}) $.}

	\subsubsection{Channel Estimation} 
	{It is essential for the BS to obtain the CSI for hybrid beamforming.} Here we consider estimation of the low-dimensional equivalent CSI. Thanks to channel reciprocity in TDD systems, we only need to estimate the uplink channels. Thus, we consider an uplink pilot training stage before data transmission.
	The $ k $-th user first sends training pilots 
	${\tilde{\mathbf{X}}_{eq,k}} \in {\mathbb{C}^{{N_t^{RF}} \times L}}$ to the BS, where $ L $ denotes the length of pilots. Then, the received signal at the BS $\tilde{\mathbf{Y}}_{eq,k} \in {\mathbb{C}^{{N_r^{RF}} \times L}}$ is given by
	\begin{equation}
	\tilde{\mathbf{Y}}_{eq,k}\! = \!{\mathbf{H}}_{eq,k}\tilde{\mathbf{X}}_{eq,k}\! +\! {\mathbf{W}}^{H}_{RF,k}\!\mathbf{H}_{k} \!\!\!\!\sum \limits_{u=1,u\neq k}\limits^{K}\!\!\!\! {\mathbf{F}}_{RF,u}\!\tilde{\mathbf{X}}_{eq,u}\!+\! \tilde{\mathbf{Z}}_{eq,k},\label{Y_eqk}
	\end{equation}
	where $\tilde{\mathbf{Z}}_{eq,k}$ denotes the AWGN.
	The transmitted pilot signal in the $l$-th pilot slot (the $l$-th column of $\tilde{\mathbf{X}}_{eq,k}$) should meet the power constraint: ${\|\tilde{\mathbf{x}}_{eq,k,l}\|}^2 \leq P$.
	Then, the BS estimates the channel $\hat{\mathbf{H}}_{eq,k} \in \mathbb{C}^{N_r \times N_t}$ based on the received signal $\tilde{\mathbf{Y}}_{eq,k}$ and the pilot $\tilde{\mathbf{X}}_{eq,k}$, which can be expressed as 
	\begin{equation}
	\hat{\mathbf{H}}_{eq,k} = \mathcal{F}(\tilde{\mathbf{Y}}_{eq,k}, \tilde{\mathbf{X}}_{eq,k}),\label{H_hat}
	\end{equation}
	where $ \mathcal{F}(\cdot) $ denotes a specific channel estimation algorithm.
	\subsubsection{Hybrid Beamforming}
	After acquiring the channel information, the BS designs the hybrid beamformers based on the channel $ {\mathbf{H}}_{k} $. The hybrid beamforming design scheme  $ \mathcal{Q}(\cdot) $ at the transmitter can be denoted as
	\begin{equation}
		\{\mathbf{F}_{RF,k}, \mathbf{W}_{RF,k}, \mathbf{F}_{BB,k}, \mathbf{W}_{BB,k}\} =\mathcal{Q}({\mathbf{H}}_{k}), \forall k.\label{Q(H)}	
	\end{equation}
	\subsection{Mixed-Timescale Frame Structure}

	{Generally, the dimension of the channel matrix is high in massive MIMO systems due to the large number of antennas. It is therefore impractical to estimate each instantaneous CSI due to the unacceptable signaling overhead and high computational complexity. To address this problem, we consider a practical mixed-timescale frame structure as shown in Fig. \ref{Twotime},  which takes into consideration both the instantaneous CSI and the channel statistics. We consider a superframe during which the channel statistics are constant. It consists of $ T_{f} $ frames, each of which is made of  $ T_{s} $ time slots. The instantaneous CSI remains unchanged during each time slot. We introduce two different timescales as follows:
	\begin{itemize}
		\item Long-timescale: The channel statistics are unchanged during each superframe which consists of several time slots.
		\item Short-timescale: The instantaneous CSI is constant in each time slot.
	\end{itemize}
	
	\begin{figure}[t]
	\begin{centering}
		\includegraphics[width=0.5\textwidth]{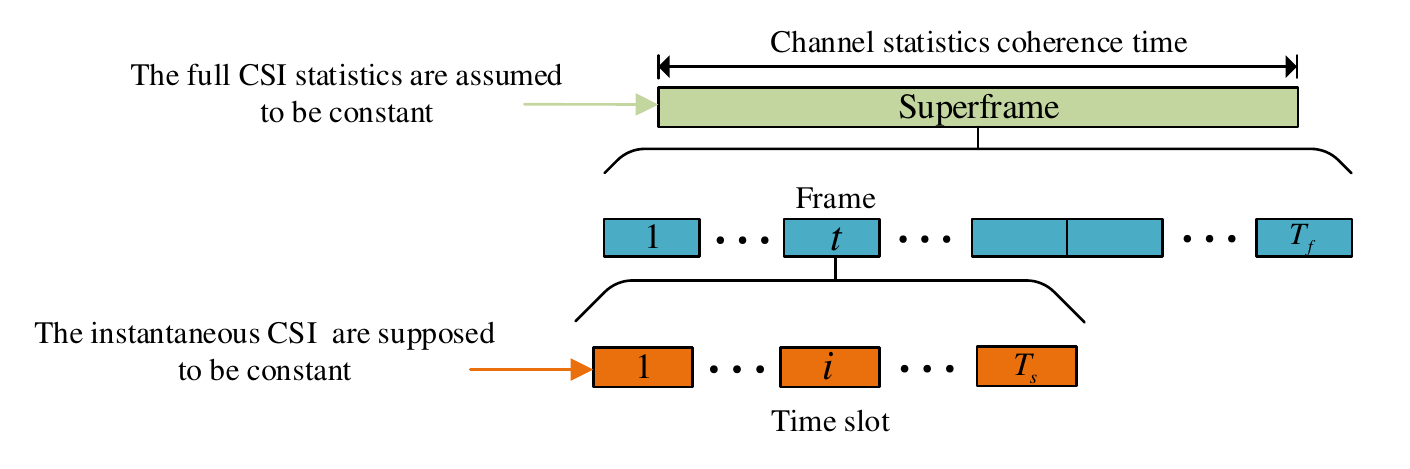}
		\par\end{centering}
	\caption{Mixed-timescale frame structure.}
	\label{Twotime}
	\end{figure}
	 In general, the dimension of the equivalent CSI matrix $ \mathbf{H}_{eq} \in \mathbb{C}^{N_{r}^{RF} \times N_{t}^{RF}} $ is much smaller than the dimension of the full CSI matrix $ \mathbf{H}\in\mathbb{C}^{N_r \times N_t}  $. Thus, we consider acquiring the low-dimensional equivalent CSI at each time slot. In this way, we can optimize the analog and digital beamformers at different timescales. We update the long-term analog beamformers $ \{\mathbf{F}_{RF}, \mathbf{W}_{RF} \} $ based on the long-term channel statistics when channel statistics \footnote{
 	 In this work, channel statistics refer to the moments or distribution of the
 	 channel fading realizations. In the mixed-timescale scheme, we need to obtain several (potentially outdated) channel samples at each superframe, and the analog beamformers are optimized based on the observed channel samples.} change, and optimize the short-term digital beamformers $ \{\mathbf{F}_{BB}, \mathbf{W}_{BB} \} $  based on the low-dimensional equivalent CSI at each time slot. 
	}
  
	\subsection{Problem Formulation}
	We aim at jointly designing the mixed-timescale hybrid beamforming and channel estimation to maximize the system sum rate. The optimization problem within each frame can be formulated as
	\begin{subequations}  \label{Problem}
		\begin{eqnarray}
		& \max \limits_{\mathcal{X}, \mathcal{Y}} & \sum \limits_{k=1}^K \sum \limits_{l=1}^{N_{s}} \log(1 + \Gamma_{k, l}),  \label{objective_function}\\
		&\text{s.t.}  &  \parallel[\mathbf{F}_{RF,k}]_{ij}\parallel ^2 = \frac{1}{\sqrt{N_t}}, \forall k,i,j ,\label{modulus1}\\
		& & \parallel[\mathbf{W}_{RF,k}]_{ij}\parallel ^2 = \frac{1}{\sqrt{N_r}},  \forall k,i,j , \label{modulus2} \\
		& & \parallel\mathbf{F}_{RF,k}\mathbf{F}^{i}_{BB,k}\parallel_{F}^2 = N_{s},  \forall k, \label{Power} \\
		& & \|\mathbf{\tilde{x}}_{eq,k,l}\|^2 \leq P, \forall k,l, \label{Pilots} \\
		& & \hat{\mathbf{H}}^{i}_{eq,k} = \mathcal{F}_{eq}(\tilde{\mathbf{Y}}_{eq,k}, \tilde{\mathbf{X}}_{eq,k}), \forall k, \label{Heq_hat1} \\
		& & \{\mathbf{F}_{RF,k}, \mathbf{W}_{RF,k}\} = \mathcal{Q}_{fu}({\mathbf{\mathbf{H}}}_{k}), \forall k, \label{Beamformers}\\
		& & \{\mathbf{F}^{i}_{BB,k}, \mathbf{W}^{i}_{BB,k}\} = \mathcal{Q}_{eq}(\hat{\mathbf{H}}^{i}_{eq,k}), \forall k, \label{Beamformers1}
		\end{eqnarray}
	\end{subequations}
    where $ \mathcal{X}  \triangleq \{\mathbf{F}_{RF,k}, \mathbf{W}_{RF,k} ,\mathbf{F}^{i}_{BB,k}, \mathbf{W}^{i}_{BB,k}\} $ and $ \mathcal{Y} \triangleq \{\tilde{\mathbf{X}}_{eq,k} \} $. In particular, $ \mathbf{\mathbf{H}}_{k} $ denotes long-term channel statistics, $ \hat{\mathbf{H}}^{i}_{eq,k} $ denotes the estimated equivalent CSI, and $ \{\mathbf{F}^{i}_{BB,k}, \mathbf{W}^{i}_{BB,k}\} $ represent the digital beamformers at the $ i $-th time slot. In addition, $ \mathcal{F}_{eq}(\cdot) $ denotes the estimation scheme for low-dimensional equivalent CSI, and $ \mathcal{Q}_{fu}(\cdot) $ and $ \mathcal{Q}_{eq}(\cdot) $ represent the analog and digital beamforming schemes, respectively. Constraints (\ref{modulus1}) and (\ref{modulus2}) reflect the unit modulus constraints for analog
    beamformers, and (\ref{Power}) and (\ref{Pilots}) represent the transmit power constraints. As we can see, the mixed-timescale problem is challenging to solve. In the following, a deep-unfolding framework is proposed for tackling this problem.
	\section{Proposed mixed-timescale Deep-Unfolding Framework}
	\label{framework}
\begin{figure*}[t]
	 	\begin{centering}
	 		\includegraphics[width=0.9\textwidth]{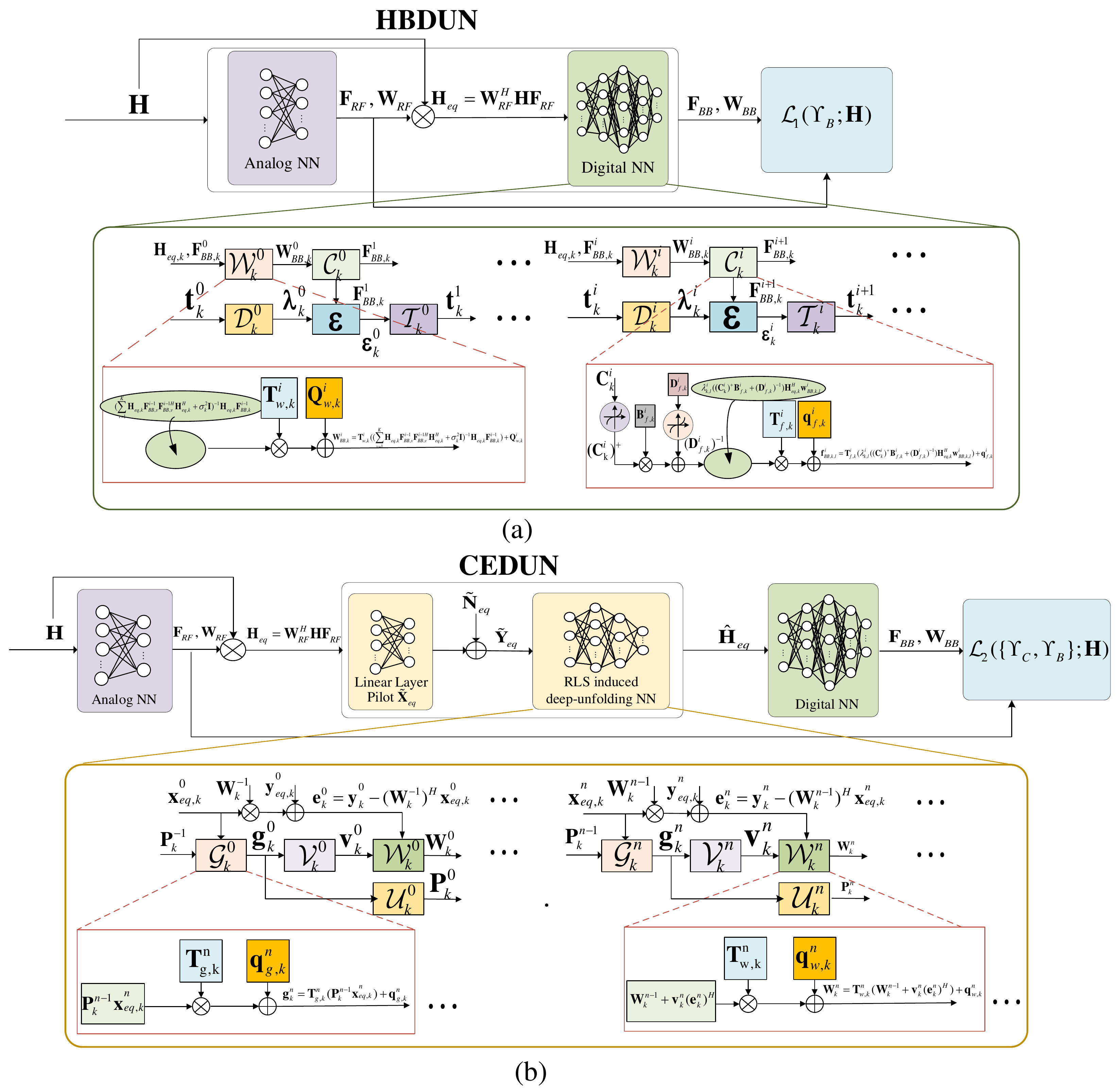}
	 		\par\end{centering}
	 	\caption{Structure of the proposed deep-unfolding framework: (a) Deep-unfolding NN for the first training stage; (b) Deep-unfolding NN for the second training stage.}
	 	\label{twotimeframework}
	 \end{figure*}

	In this section, we propose a mixed-timescale deep-unfolding framework, the structure of which is shown in Fig. \ref{twotimeframework}. We first present how to jointly train the proposed deep-unfolding NNs offline and then show the whole process of the training and data transmission under the mixed-timescale scheme.
	 \subsection{The Two-Stage Joint Training}\label{twostage}
	  {During data transmission, the analog beamformers are fixed and only digital beamformers are updated based on the estimated equivalent channel at each time slot. To model the transmission process, we divide the training process for the deep-unfolding NN into two stages where the analog beamformers are optimized in the first stage and then fixed in the second stage.}
	 \subsubsection{The First Training Stage}
	 Fig. \ref{twotimeframework}(a) denotes the deep-unfolding NN for the first training stage, which presents the architecture of the HBDUN that consists of the analog NN and digital NN for designing the analog and digital beamforming, respectively. The input of the analog NN is the full CSI sample matrix $ \mathbf{H} $  (possibly outdated) and the outputs are analog beamformers $ \{\mathbf{W}_{RF}, \mathbf{F}_{RF}\} $. The input of the digital NN is the real-time low-dimensional equivalent CSI matrix $ \mathbf{H}_{eq} $ and the outputs are digital beamformers $ \{\mathbf{W}_{BB}, \mathbf{F}_{BB}\} $.
	 \subsubsection*{(a) Forward Propagation}
	We obtain channel samples offline based on the long-term channel statistics as the input of the analog NN and the outputs are analog beamformers $ \{\mathbf{W}_{RF}, \mathbf{F}_{RF}\} $. Then we obtain the low-dimensional equivalent CSI matrices $ \mathbf{H}_{eq} $ which pass through the digital NN that outputs the digital beamformers $ \{\mathbf{F}_{BB}, \mathbf{W}_{BB}\} $. We introduce the detailed structure of the HBDUN in Section \ref{HBDUN}. The
	forward propagation for the first stage $ \mathcal{P}_{1}(\cdot) $ is expressed as 
	\begin{equation}
		\{\mathbf{W}_{RF},\mathbf{F}_{RF}, \mathbf{W}_{BB}, \mathbf{F}_{BB}\} = \mathcal{P}_{1}(\mathbf{\Upsilon}_{B};\mathbf{H}),
	\end{equation}
	where $ \mathbf{\Upsilon}_{B} $ represents the trainable parameters of the HBDUN. Note that $ \mathbf{\Upsilon}_{B} $ consists of $ \mathbf{\Psi} $ and $ \mathbf{\Omega} $, which represent the trainable parameters of the analog NN and digital NN, respectively.
	 \subsubsection*{(b) Loss Function}
	 The loss function of the first stage is denoted as $ \mathcal{L}_{1}(\mathbf{\Upsilon}_{B};\mathbf{H}) $, which is the system sum rate:
	 \begin{equation}
      \mathcal{L}_{1}(\mathbf{\Upsilon}_{B};\mathbf{H}) = - \sum \limits_{k=1}^K \sum \limits_{l=1}^{N_{s}} \log(1 + \Gamma_{k, l}).
	 \end{equation}
	 \subsubsection*{(c) Back Propagation}
	All the trainable parameters of the HBDUN are updated based on the stochastic gradient descent (SGD) algorithm. Specifically, in the $ i $-th round of the training process, we update the trainable parameters as follows:
	\begin{equation}
	\mathbf{\Upsilon}_{B}^{i+1} = \mathbf{\Upsilon}_{B}^{i} - \eta\dfrac{\partial \mathcal{L}_{1}(\mathbf{\Upsilon}_{B};\mathbf{H})}{\partial \mathbf{\Upsilon}_{B}},
	\end{equation}
	 where $ \eta $ is the learning rate.
	 \subsubsection{The Second Training Stage}
	 {Fig. \ref{twotimeframework}(b) shows the deep-unfolding NN for the second training stage which consists of the modules of channel estimation for low-dimensional equivalent CSI and digital beamforming. Note that the CEDUN represents the NN for equivalent CSI estimation, which consists of the pilot training NN and the RLS induced deep-unfolding NN. The analog and digital NN are employed to obtain analog and digital beamformers, respectively. The input of the CEDUN is the low-dimensional equivalent CSI matrix $ \mathbf{H}_{eq} $ and the output is the estimated channel matrix $ \hat{\mathbf{H}}_{eq} $.
	 \subsubsection*{(a) Forward Propagation}
	 First, we fix the trained analog NN to obtain analog beamformers $ \{\mathbf{W}_{RF}, \mathbf{F}_{RF}\} $ and the low-dimensional equivalent CSI matrix $ \mathbf{H}_{eq} $. Then $ \mathbf{H}_{eq} $ passes through the pilot training NN and the RLS deep-unfolding NN which outputs the estimated equivalent CSI matrix $ \hat{\mathbf{H}}_{eq} $. Finally, $ \hat{\mathbf{H}}_{eq} $ passes through the digital NN that outputs the digital beamformers $ \{\mathbf{F}_{BB}, \mathbf{W}_{BB}\} $. We introduce the detailed structure of the CEDUN in Section \ref{CEDUN11}.
	 The forward propagation for the second stage $ \mathcal{P}_{2}(\cdot) $ is expressed as 
	 \begin{equation}
	 \{\mathbf{W}_{RF},\mathbf{F}_{RF}, \mathbf{W}_{BB}, \mathbf{F}_{BB}\} = \mathcal{P}_{2}(\{\mathbf{\Upsilon}_{C}, \mathbf{\Upsilon}_{B}\};\mathbf{H}),
	 \end{equation}
	 where $ \mathbf{\Upsilon}_{C} $ consists of $ \tilde{\mathbf{X}}_{eq} $ and $ \mathbf{\Xi} $,  which represent the trainable parameters of the pilot training NN and the RLS induced deep-unfolding NN, respectively.}
	 \subsubsection*{(b) Loss Function}
	 The loss function of the second stage is denoted as  $ \mathcal{L}_{2}(\{\mathbf{\Upsilon}_{C}, \mathbf{\Upsilon}_{B}\};\mathbf{H}) $, which is the system sum rate.
	 \begin{equation}
	 \mathcal{L}_{2}(\{\mathbf{\Upsilon}_{C}, \mathbf{\Upsilon}_{B}\};\mathbf{H}) = - \sum \limits_{k=1}^K \sum \limits_{l=1}^{N_{s}} \log(1 + \Gamma_{k, l}).
	 \end{equation}
	 \subsubsection*{(c) Back Propagation}
	The trainable parameters of the CEDUN and digital NN are updated based on the SGD algorithm. Note that the parameters of the analog NN are not updated in the second training stage. Thus, in the $ i $-th round of the training process, we update the trainable parameters as follows:
	\begin{equation}
	\mathbf{\Upsilon}_{C}^{i+1} = \mathbf{\Upsilon}_{C}^{i} - \eta\dfrac{\partial \mathcal{L}_{2}(\{\mathbf{\Upsilon}_{C}, \mathbf{\Upsilon}_{B}\};\mathbf{H})}{\partial \mathbf{\Upsilon}_{C}},
	\end{equation}
	\begin{equation}
	\mathbf{\Omega}^{i+1} = \mathbf{\Omega}^{i} - \eta\dfrac{\partial \mathcal{L}_{2}(\{\mathbf{\Upsilon}_{C}, \mathbf{\Upsilon}_{B}\};\mathbf{H})}{\partial \mathbf{\Omega}}.
	\end{equation}
	\subsubsection{The Advantages of Joint Training}
     Compared with conventional separate designs, the proposed joint design framework has potential performance gains. Conventional algorithms optimize the channel estimation and hybrid beamforming modules separately under different objective functions, i.e., the minimization of MSE for channel estimation and maximization of sum rate for hybrid beamforming, respectively. In comparison, the proposed joint design NN ties the two modules tightly and jointly trains them. It provides an end-to-end deep-unfolding framework without explicitly estimating the CSI and both of the channel estimation and hybrid beamforming are designed to maximize the sum rate, which leads to better sum rate performance and is more efficient in terms of the signaling overhead.
     \subsection{The Training and the Data Transmission Stage}
\begin{figure}[t]
		\begin{centering}
			\includegraphics[width=0.5\textwidth]{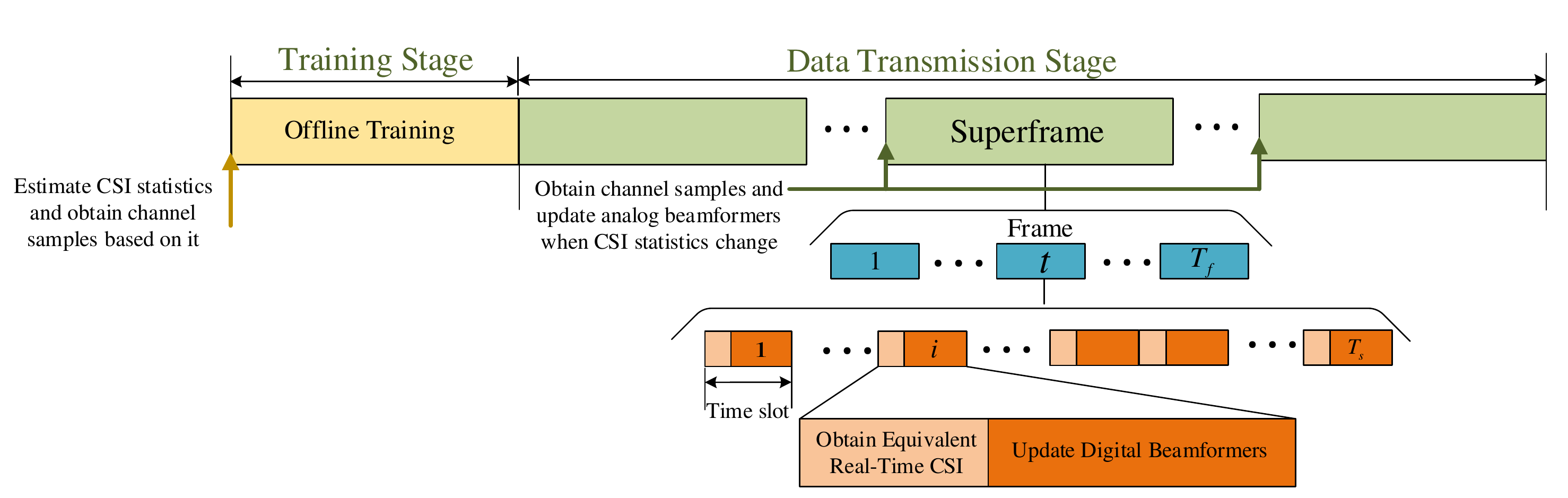}
			\par\end{centering}
		\caption{Training stage and data transmission of the offline mixed-timescale scheme.}
		\label{mixed-timescale scheme}
	\end{figure}

     Training and data transmission for the mixed-timescale is shown in Fig. \ref{mixed-timescale scheme}.
     We first obtain channel samples based on the long-term channel statistics offline and jointly train the deep-unfolding NN as mentioned above. Then in the data transmission stage, we fix the analog beamformers during the channel statistics coherence time and estimate the low-dimensional equivalent real-time CSI by CEDUN to update the digital beamformers. The analog beamformers obtained by offline training can well adapt to CSI statistics when it does not change \cite{Twotime1}. When it changes, we obtain channel samples and employ transfer learning \cite{transferlearning} to fine tune the parameters of the deep-unfolding NN, where the analog beamformers are updated to better fit the change of CSI statistics \cite{CSI}.
	\section{Deep-unfolding Network for Channel Estimation}
	\label{Section3:CE}
	In this section, we first introduce a RLS based algorithm for estimating the low-dimensional equivalent CSI matrices and then  describe the deep-unfolding NN by introducing trainable parameters.
	\subsection{The RLS Channel Estimation Algorithm}
    We focus on the expression (\ref{Y_eqk}), which represents the received pilot signal for user $ k $. We aim to obtain the estimated low-dimensional equivalent channel matrix $ \hat{\mathbf{H}}_{eq,k} $ by minimizing the mean square estimation error which is defined as $ \mathbb{E}\{{\| \hat{\mathbf{H}}_{eq,k} - \mathbf{H}_{eq,k}\|^{2}}\} $. The solution of the least squares (LS) approach for solving this problem is given by
    \begin{equation}
    \hat{\mathbf{H}}_{eq,k} = \tilde{\mathbf{Y}}_{eq,k} {\tilde{\mathbf{X}}_{eq,k}}^{H} (\tilde{\mathbf{X}}_{eq,k}{\tilde{\mathbf{X}}_{eq,k}}^{H})^{-1}.\label{LS}
    \end{equation}
    In the RLS algorithm, the matrix inversion in (\ref{LS}) is replaced by an iterative process. The procedure of the RLS channel estimation algorithm is presented in Algorithm 1, where
    $ \tilde{\mathbf{x}}_{eq,k}^{n} $ and $ \tilde{\mathbf{y}}_{eq,k}^{n} $ are the $ n $-th column of $ \tilde{\mathbf{X}}_{eq,k} $ and $ \tilde{\mathbf{Y}}_{eq,k} $, respectively. Note that $ \mathbf{g}_{k}^{n}, \mathbf{v}_{k}^{n} $, and $\mathbf{P}_{k}^{n}$ are intermediate variables, $ \beta_{k} \in (0, 1)$ is the forgetting factor and  $ \delta $ denotes a small positive number, $ \mathbf{W}_{k}^{n} $ denotes the weight matrix and the estimated channel $ \hat{\mathbf{H}}_{eq,k} = (\mathbf{W}_{k}^{n})^{H}$. Moreover, the estimation error $ \mathbf{e}_{k}^{n} $ descends with the update of $ \mathbf{W}_{k}^{n} $. The number of iterations is the length of pilots $ L $. In addition, the inputs of the algorithm are the pilots and received signal, and the output is the estimated channel matrix.
	
   	\begin{algorithm}[t] 
	\caption{The RLS algorithm for channel estimation} 
	\begin{algorithmic}[1]
		\begin{small}
			\State Input: Pilot $ \tilde{\mathbf{X}}_{eq,k} $ and received signal $ \tilde{\mathbf{Y}}_{eq,k} $;
			\State Initialize the estimated matrix $\mathbf{W}_{k}^{-1} = \mathbf{0}$ and 	intermediate variable $\mathbf{P}_{k}^{-1} = \delta^{-1}\mathbf{I}$;
			\For{$ n = 1, 2, \ldots, L$}
			\State Update $\{\mathbf{g}_{k}^{n}\}$ based on $\mathbf{g}_{k}^{n} = \mathbf{P}_{k}^{n-1}\tilde{\mathbf{x}}_{eq,k}^{n-1}$;
			\State Update $\{\mathbf{v}_{k}^{n}\}$ based on $\mathbf{v}_{k}^{n} = \frac{\mathbf{g}_{k}^{n}}{\beta_{k} + (\mathbf{g}_{k}^{n})^{H}\tilde{\mathbf{x}}_{eq,k}^{n}}$;
			\State Update $\{\mathbf{P}_{k}^{n}\}$ based on $\mathbf{P}_{k}^{n} = \beta_{k}^{-1}(\mathbf{P}_{k}^{n-1} - \mathbf{v}_{k}^{n}(\mathbf{g}_{k}^{n})^{H})$;
			\State Calculate residual $\mathbf{e}_{k}^{n} = \tilde{\mathbf{y}}_{k}^{n} - (\mathbf{W}_{k}^{n-1})^{H}\tilde{\mathbf{x}}_{eq,k}^{n}$;
			\State Update the estimated matrix $\mathbf{W}_{k}^{n} = \mathbf{W}_{k}^{n-1} + \mathbf{v}_{k}^{n}(\mathbf{e}_{k}^{n})^{H}$; 
			\State $  n = n + 1 $;
			\EndFor
		\end{small}
	\end{algorithmic}
\label{rls}
	\end{algorithm}
	\subsection{Deep-Unfolding NN for Channel Estimation}\label{CEDUN11}
	Based on the RLS algorithm, we propose the CEDUN which contains the pilot training NN and RLS induced deep-unfolding NN. In the proposed offline mixed-timescale framework, we only need to estimate the low-dimensional equivalent CSI at each time slot. To estimate the equivalent channel $ \mathbf{H}_{eq,k} $, the $k$-th user sends the training pilot matrix $ \tilde{\mathbf{X}}_{eq,k} $, and at the BS the received pilot signal matrix $ \tilde{\mathbf{Y}}_{eq,k} $ is denoted as (\ref{Y_eqk}). Then $ \tilde{\mathbf{X}}_{eq,k} $ and $ \tilde{\mathbf{Y}}_{eq,k} $ are input to the RLS induced deep-unfolding NN to obtain the estimated equivalent channel $\hat{\mathbf{H}}_{eq,k}$. The structure of the CEDUN for user $ k $ is shown in Fig. \ref{twotimeframework}(b). 
	\subsubsection{Pilot Training NN}
	Different from the conventional Gaussian pilots and discrete fourier transform (DFT) pilots, in the proposed deep-unfolding NN, we set the pilots as trainable parameters that can adapt to the CSI statistics to further improve the performance. As shown in Fig. \ref{twotimeframework}(b), to model the process of pilot training for estimating the low-dimensional equivalent CSI matrix $ \mathbf{H}_{eq,k} $, the input and output of the NN are $ \mathbf{H}_{eq,k} $ and $ \tilde{\mathbf{Y}}_{eq,k} $, respectively, and we set $ \tilde{\mathbf{X}}_{eq} $ as the trainable parameter. Note that $ \tilde{\mathbf{X}}_{eq} $ needs to be scaled to satisfy the power constraint (\ref{Pilots}). 
	\subsubsection{RLS Induced Deep-Unfolding NN}
	We unfold the RLS into a network with significantly less layers. The inputs of the $ n $-th layer of the NN are $ \{\tilde{\mathbf{x}}_{eq,k}^{n}, \tilde{\mathbf{y}}_{eq,k}^{n}, \mathbf{P}_{k}^{n-1}, \mathbf{W}_{k}^{n-1} \}$ and the outputs are $ \{\mathbf{P}_{k}^{n}, \mathbf{W}_{k}^{n} \}$.

    {
    To increase the degrees of freedom, we introduce the structure 
    \begin{equation}
    \mathbf{Y}_{out}^{n} = \mathbf{T}_{y}^{n} \mathbf{X}_{in}^{n} + \mathbf{q}_{y}^{n},
    \end{equation}
    where $ \mathbf{X}_{in}^{n} $ and $ \mathbf{Y}_{out}^{n} $ represent the input and output of the $ n $-th layer, respectively, $ \mathbf{T}_{y}^{n} $ and $ \mathbf{q}_{y}^{n} $ are the introduced multiplier and offset trainable parameter of the $ n $-th layer, respectively.
    Note that $ \varpi^{n}_{C} \triangleq \{\mathbf{T}_{g,k}^{n}, \mathbf{q}_{g,k}^{n}\} \cup \{ \mathbf{T}_{v,k}^{n}, \mathbf{q}_{v,k}^{n}\} \cup \{ \mathbf{T}_{p,k}^{n}, \mathbf{q}_{p,k}^{n}\} \cup \{ \mathbf{T}_{w,k}^{n}, \mathbf{q}_{w,k}^{n}\} $ are the multiplier and offset trainable parameters to update the variables $ \mathbf{g}_{k}^{n} $, $ \mathbf{v}_{k}^{n} $, $ \mathbf{P}_{k}^{n} $, and $ \mathbf{W}_{k}^{n} $ in the $ n $-th layer, respectively. 
    As shown in Fig. \ref{twotimeframework}(b), based on Algorithm \ref{rls}, $ \mathcal{G}_{k}^{n} $, $ \mathcal{V}_{k}^{n} $, $ \mathcal{U}_{k}^{n} $, $ \mathcal{W}_{k}^{n} $ represent the sub-layers of the $ n $-th layer of the deep-unfolding NN, i.e., (\ref{g^n,k})-(\ref{W^n_k}). 
}
  	 \begin{subequations}
    	\begin{eqnarray}
    	&&{\mathbf{g}}_{k}^{n} = \mathbf{T}^{n}_{g,k} (\mathbf{P}_{k}^{n-1} \tilde{\mathbf{x}}_{eq,k}^{n}) + \mathbf{q}_{g,k}^{n},\label{g^n,k}\\
    	&&{\mathbf{v}}_{k}^{n} = \mathbf{T}^{n}_{v,k} (\frac{{\mathbf{g}}_{k}^{n}}{\gamma_{k}^{n} + {({\mathbf{g}}_{k}^{n})^{H}} \tilde{\mathbf{x}}_{eq,k}^{n}}) + \mathbf{q}_{v,k}^{n},\label{v^n,k}\\
    	&&{\mathbf{P}}_{k}^{n} \!=\!\mathbf{T}^{n}_{p,k}(\gamma_{k}^{n})^{-1}(\mathbf{P}_{k}^{n-1} - {\mathbf{v}}_{k}^{n} ({\mathbf{g}}_{k}^{n})^{H}) \!+ \!\mathbf{q}_{p,k}^{n}, \label{P^n_k}\\
    	&&{\mathbf{W}}_{k}^{n} = \mathbf{T}_{w,k}^{n} (\mathbf{W}_{k}^{n-1} + \mathbf{v}_{k}^{n}(\mathbf{e}_{k}^{n})^{H}) + \mathbf{q}_{w,k}^{n}.\label{W^n_k}
    	\end{eqnarray}
    \end{subequations}
{
	Thus, the trainable parameters of the deep-unfolding NN are $ \mathbf{\Xi} \triangleq \bigcup_{n=1}^{L_{c}} \varpi^{n}_{C}\cup \gamma_{k}^{n} $, where $ L_{c} $ is the number of layers. 
	All the trainable parameters of the CEDUN are denoted as $ \mathbf{\Upsilon}_{C} \triangleq \tilde{\mathbf{X}}_{eq} \cup \mathbf{\Xi} $.}

    \section{Deep-unfolding Network for Hybrid Beamforming}
    \label{HB}
    {
    We next turn to design an SSCA algorithm for hybrid beamforming and then introduce the proposed HBDUN which unfolds the SSCA algorithm.
	\subsection{The SSCA-based Hybrid Beamforming Algorithm}}
{
	For the hybrid beamforming design, we first introduce the mixed-timescale SSCA framework \cite{Twotime1}. Let us express the element of analog beamformers in terms of $ {e}^{j\phi} $ as
	\begin{subequations}
    	\makeatletter
		\def\@eqnnum{{\normalfont  (\theequation)}} \makeatother
		\begin{eqnarray}
			&&\mathbf{F}_{RF} = e^{j\bm{\phi}_{F}}, \bm{\phi}_{F}\in\mathbb{C}^{N_{t}\times  N_{t}^{RF}},\label{shifterF}\\
			&&\mathbf{W}_{RF} = e^{j\bm{\phi}_{W}}, \bm{\phi}_{W}\in\mathbb{C}^{N_{r}\times  N_{r}^{RF}}, \label{shifterW}
		\end{eqnarray}
	\end{subequations}
}
{
	\hspace{-2.2mm}where $ \bm{\phi} \triangleq \{\bm{\phi}_{F}, \bm{\phi}_{W}\} $ denotes the angle of phase shifter. Here, the objective function can be expressed as $ r_{0}(\bm{\phi}, \mathbf{M}; \mathbf{H}) $, where  $ \mathbf{M} \triangleq \{\mathbf{F}_{BB}, \mathbf{W}_{BB}\} $.
\subsubsection{Long-Term Analog Beamformers}
	We optimize the analog beamformers based on the full CSI samples $ \mathbf{H} $. First, we fix the digital beamformers $ \mathbf{M}^{J} $, which means that the digital beamformers are obtained by running the algorithm that optimizes the short-term variables for $ J $ iterations. Then we employ a convex surrogate function to replace the objective function to optimize the analog beamformers. The surrogate function in the $ t $-th iteration is\cite{Twotime1}
	\begin{equation}
		\bar{f}^{t}(\bm{\phi}) = f^{t} + (\bm{f}^{t}_{\phi})^{T}(\bm{\phi} - \bm{\phi}^{t}) + \tau\parallel\bm{\phi} - \bm{\phi}^{t}\parallel^{2},\label{surrogate}
	\end{equation}
	where
	\begin{equation}
		\bm{f}^{t}_{\phi} = (1 - \rho^{t})\bm{f}^{t-1}_{\phi} + \rho^{t}\bigtriangledown_{\phi}r_{0}(\bm{\phi}^{t}, \mathbf{M}^{J}; \mathbf{H})
	\end{equation}
	with $ \bm{f}^{-1}_{\phi} = \bm{0} $. Here $ \bigtriangledown_{\phi}r_{0}(\bm{\phi}^{t} , \mathbf{M}^{J};\mathbf{H})$ is the partial derivative of $ r_{0}(\bm{\phi}^{t}, \mathbf{M}^{J}; \mathbf{H}) $,
	and the constant $ {f}^{t} $ is calculated as
	\begin{equation}
	f^{t} = (1 - \rho^{t})f^{t-1} + \rho^{t}r_{0}(\bm{\phi}^{t}, \mathbf{M}^{J}; \mathbf{H})
	\end{equation}
	with $ f^{-1} = 0 $.
	We take the derivative with formula (\ref{surrogate}) to update the long-term variable $ \bm{\phi} $ as
	\begin{equation}
		\bar{\bm{\phi}} = \bm{\phi} - \eta\bm{f}^{t}_{\phi},\label{update}
	\end{equation}
	where $ \eta $ denotes the step size. 
	   	\begin{algorithm}[t]
		
		\caption{The SCA algorithm for digital beamforming} 
		\begin{algorithmic}[1]
			\begin{small}
				\State Input: The equivalent channel $ \mathbf{H}_{eq} $.
				\State Initialize $\mathbf{t}_{k,l} = \mathbf{0}$ and digital precoder $\mathbf{F}_{BB}$ satisfying the power constraint condition and set the maximum iteration number $ I_{max} $;
				\For{$ i = 1, 2, 3, \ldots$}
				\State Update digital combiner $ \mathbf{W}_{BB,k}^{i} $ based on (\ref{W_{BB}});
				\State Update dual variable $ \lambda_{k,l}^{i} = \frac{\alpha^{\mathbf{t}^{i}_{k,l}}}{\log{\alpha}}$;
				\State Update digital precoder
				\begin{equation}
				\begin{aligned}
				\mathbf{f}_{BB,k,l}^{i} & = \lambda_{k,l}^{i}\Bigg(\sum\limits_{(m,n)}^{(K,N_{s})} \lambda_{m,n}^{i} \mathbf{H}_{eq,m}^{H} \mathbf{w}_{BB,m,n}^{i}\\ &(\mathbf{w}_{BB,m,n}^{i})^{H}\mathbf{H}_{eq,m} + \tau_k \mathbf{I} \Bigg)^{-1} \mathbf{H}_{eq,k}^{H} \mathbf{w}_{BB,k,l}^{i};
				\end{aligned} 
				\nonumber
				\end{equation}
				\State Calculate MSE $\epsilon_{k,l}^{i} $ based on (\ref{mse});
				\State Update $\mathbf{t}_{k,l}^{i+1}  = \mathbf{t}_{k,l}^{i} + \frac{1}{\log{\alpha}}(1 - \epsilon_{k,l}^{i}\alpha^{\mathbf{t}_{k,l}^{i}})$;\\
				\textbf{Until} desired level of convergence or $ i  >  I_{max} $.
				\EndFor
			\end{small}
		\end{algorithmic}
		\label{sca}
	\end{algorithm}
	\subsubsection{Short-Term Digital Beamformers}
	By fixing the analog beamformers, we optimize the digital beamformers based on the low-dimensional equivalent CSI matrix $ \mathbf{H}_{eq} $. We adopt the successive convex approximation (SCA) algorithm to optimize the short-term digital beamformers \cite{SCA}. For the data stream $ l $ of user $ k $, the MSE expression is given as
	\begin{equation}
	\begin{aligned}
		\mathbf{\epsilon}_{k,l} &= |1 - \mathbf{w}_{BB,k,l}^{H}\mathbf{H}_{eq,k}\mathbf{f}_{BB,k,l}|^{2}\\
		& + \!\!\!\!\!\!\!\!\sum\limits^{K,N_{s}}\limits_{m,n(m \neq k,n \neq l)}\!\!\!\!\!\!\!\!\!|\mathbf{w}_{BB,k,l}^{H}\mathbf{H}_{eq,k}\mathbf{f}_{BB,m,n}|^{2} 
		\!\!+ \!\!\mathbf{\sigma}_{k}^{2}\parallel \!\!\mathbf{w}_{BB,k,l}\!\!\parallel^{2}. \label{mse}
	\end{aligned}
	\end{equation}
	The optimal digital beamformer is given by
	\begin{equation}
	\mathbf{W}_{BB,k} \!\!=\!\! (\sum\limits_{v=1}\limits^{K}\mathbf{H}_{eq,k}\!\mathbf{F}_{BB,v}\!\mathbf{F}_{BB,v}^{H}\!\mathbf{H}_{eq,k}^{H} \!+\! \sigma_{k}^2\mathbf{I})^{-1}\!\mathbf{H}_{eq,k}\!\mathbf{F}_{BB,k}.\label{W_{BB}}
	\end{equation}
	We equivalently transform the optimization objective function into the problem that minimizes the MSE \cite{SCA}:
	\begin{equation}
		\min \limits_{\mathbf{F}_{BB}}\sum\limits_{k=1}\limits^K\sum\limits_{l=1}\limits^{N_{s}}\log(\mathbf{\epsilon}_{k,l}). \label{minlog}
	\end{equation}
	Then we design the digital precoder $ \mathbf{F}_{BB} $. By introducing a monotonic log-concave function $ g(\mathbf{t}_{k,l}) $, the target optimization function becomes \cite{SCA}
	\begin{subequations}
		\label{g}
    	\makeatletter
		\def\@eqnnum{{\normalfont (\theequation)}} \makeatother
		\begin{eqnarray}
			&\min\limits_{\mathbf{F}_{BB}, \mathbf{t}_{k,l}}&\sum\limits_{k=1}\limits^K\sum\limits_{l=1}\limits^{N_{s}}\log(g(\mathbf{t}_{k,l})^{-1}),\\
		    &\text{s.t.}
		    &\mathbf{\epsilon}_{k,l}\leq g(\mathbf{t}_{k,l})^{-1}. 
		\end{eqnarray}
	\end{subequations}
}
{
    \hspace{-2mm}We then use the first-order Taylor approximation
    \begin{equation}
    	\bar{g}(\mathbf{t}_{k,l},\mathbf{t}_{k,l}^{0}) = g(\mathbf{t}_{k,l}^{0}) + (\mathbf{t}_{k,l} - \mathbf{t}_{k,l}^{0})\frac{\partial}{\partial \mathbf{t}_{k,l}}g(\mathbf{t}_{k,l}^{0}).
    \end{equation}
    Problem (\ref{g}) can be formulated as
    \begin{subequations}
    	\makeatletter
    	\def\@eqnnum{{\normalfont (\theequation)}} \makeatother
    	\begin{eqnarray}
    	&\min\limits_{\mathbf{F}_{BB}, \mathbf{t}_{k,l}}&\sum\limits_{k=1}\limits^K\sum\limits_{l=1}\limits^{N_{s}}\log(\bar{g}(\mathbf{t}_{k,l}, \mathbf{t}_{k,l}^{(i)})),\\
    	&\text{s.t.}
    	&\mathbf{\epsilon}_{k,l}\leq[\bar{g}(\mathbf{t}_{k,l}, \mathbf{t}_{k,l}^{(i)})].
    	\end{eqnarray}
    \end{subequations}
     To simplify the problem, we consider a log-linear function $ g(x) = \alpha^{x} $, where $ \alpha > 1 $. We present the SCA algorithm in Algorithm \ref{sca}, where $ \tau_k $ denotes the Lagrange multiplier.
    }
	\subsection{The SSCA Algorithm Induced Deep-Unfolding NN} \label{HBDUN}
	We next unfold the SSCA algorithm leading to HBDUN. The structure of HBDUN is shown 
	in Fig. \ref{twotimeframework}(a), where the first network and the second network are referred to as the analog and digital NNs, respectively.
	The input of the analog NN is the full channel samples $ \mathbf{H} $ and the outputs are 
	the analog beamforming matrix $ \{\mathbf{F}_{RF,k}, \mathbf{W}_{RF,k}\} $. We set the angle of the phase shifters for analog beamformers as trainable parameters $ \mathbf{\Psi} \triangleq \{\mathbf{\Psi}_{F}, \mathbf{\Psi}_{W}\} $ of the analog NN and employ the operation $ e^{j(\cdot)} $ to satisfy the unit modulus constraint.

	 The input of the digital NN is the low-dimensional real-time equivalent CSI matrix $ \mathbf{H}_{eq,k} $ and the outputs are the digital beamforming matrix $ \{\mathbf{F}_{BB,k}, \mathbf{W}_{BB,k}\} $. We next introduce the detailed structure of the digital NN, which unfolds the SCA algorithm into a layer-wise structure. {Two non-linear operations are defined for approximating matrix inversion: (i) We take the inverse of the diagonal entries of matrix $ \mathbf{A} $ and set the other elements in $ \mathbf{A} $ as zero, and denote the result as $ \mathbf{A}^{+} $. For example,
	 \begin{equation}
	 \mathbf{A} = \begin{bmatrix}
	 a_{11} & a_{12} & a_{13} \\
	 a_{21} & a_{22} & a_{23} \\
	 a_{31} & a_{32} & a_{33}
	 \end{bmatrix},
	 \mathbf{A}^{+} = \begin{bmatrix}
	 {a_{11}^{-1}} & 0 & 0 \\
	 0 & {a_{22}^{-1}} & 0 \\
	 0 & 0 & {a_{33}^{-1}}
	 \end{bmatrix};
	 \end{equation} (ii) we set the imaginary part of the diagonal elements of $ \mathbf{D} $ to zero, expressed as $ \mathbf{D}^{-} $:
	 \begin{equation}
	 \begin{aligned}
	 &&\mathbf{D} \!=\! \begin{bmatrix}
	 d_{r,11}\!+\!jd_{i,11} \!&\! d_{r,12}\!+\!jd_{i,12} \!&\! d_{r,13}\!+\!jd_{i,13} \\
	 d_{r,21}\!+\!jd_{i,21} \!&\! d_{r,22}\!+\!jd_{i,22} \!&\! d_{r,23}\!+\!jd_{i,23} \\
	 d_{r,31}\!+\!jd_{i,31} \!&\! d_{r,32}\!+\!jd_{i,32} \!&\! d_{r,33}\!+\!jd_{i,33}
	 \end{bmatrix}\!, \\
	 &&\!\mathbf{D}^{-} \!=\!\! \begin{bmatrix}
	 \!\!\! d_{r,11} \!\!&\!\! d_{r,12}\!+\!jd_{i,12} \!&\! d_{r,13}\!+\!jd_{i,13} \\
	 d_{r,21}\!+\!jd_{i,21} \!&\! d_{r,22} \!&\! d_{r,23}\!+\!jd_{i,23} \\
	 d_{r,31}\!+\!jd_{i,31} \!&\! d_{r,32}\!+\!jd_{i,32} \!&\! d_{r,33}
	 \end{bmatrix}.
	 \end{aligned}
	 \end{equation}
	 Based on this, we employ the following structure to approximate the matrix inversion.
	 \begin{itemize}
	 	\item First, we use $ \mathbf{A}^{+} \mathbf{B} $ with non-linear operation $ \mathbf{A}^{+} $ and trainable parameter $ \mathbf{B} $, where $ \mathbf{B} $ is introduced to improve performance. It can be seen that $ \mathbf{A}^{-1} = \mathbf{A}^{+} $ if matrix $ \mathbf{A} $ is diagonal. We observe that the diagonal elements of the matrix are much larger than the other elements in the SCA algorithm. 
	 	\item Secondly, we introduce the offset trainable matrix $ \mathbf{D} $ to better approximate the inverse matrix. We find that the imaginary part of the diagonal elements of the inverse matrix are close to zero. Thus, we employ $ \mathbf{D}^{-} $ as the offset. 
	 \end{itemize}
	 Thus, the matrix inversion $ \mathbf{A}^{-1} $  is approximated by $ \mathbf{A}^{+}\mathbf{B} + \mathbf{D}^{-} $. The computational complexity of matrix inversion is $ \mathcal{O}(n^3) $ while that of the approximation is $ \mathcal{O}(n^{2.37}) $.}  Note that we introduce trainable parameters $ \{ \mathbf{B}_{f,k}^{i}, \mathbf{D}_{f,k}^{i}\} $
    to approximate the inversion of variable $ \mathbf{f}_{BB,k,l}^{i} $ in the $ i $-th layer, which reduces the computational complexity. To increase the
    degrees of freedom for the parameters, the multiplier and offset trainable parameters $ \bm{\varpi}^{i}_{B} \triangleq  \{ \mathbf{T}_{w,k}^{i}, \mathbf{Q}_{w,k}^{i}\} \cup \{ \mathbf{T}_{\lambda,k}^{i}, \mathbf{q}_{\lambda,k}^{i}\} \cup \{\mathbf{T}_{f,k}^{i}, \mathbf{q}_{f,k}^{i}\} \cup \{ \mathbf{T}_{t,k}^{i}, \mathbf{q}_{t,k}^{i}\} $  are introduced in updating the variables $ \mathbf{W}_{BB,k}^{i} $, $ \mathbf{\lambda}_{k,l}^{i} $, $ \mathbf{f}_{BB,k,l}^{i} $, and $ \mathbf{t}_{k,l}^{i} $ in the $ i $-th layer, respectively.
    As shown in Fig. \ref{twotimeframework}(a), based on Algorithm \ref{sca}, $ \mathcal{W}_{k}^{i} $, $ \mathcal{D}_{k}^{i} $, $ \mathcal{C}_{k}^{i} $, and $ \mathcal{T}_{k}^{i} $ represent the sub-layers of the $ i $-th layer of the deep-unfolding NN, i.e., (\ref{W^i,k})-(\ref{t^i_k}), where 
    $ \mathbf{C}^{i}_{k} \triangleq \sum\limits_{(m,n)}^{(K,N_{s})} \lambda_{m,n}^{i} \mathbf{H}_{eq,m}^{H} \mathbf{w}_{BB,m,n}^{i} (\mathbf{w}_{BB,m,n}^{i})^{H} \mathbf{H}_{eq,m} + v_{k}^{i} \mathbf{I} $. In addition, the constant $ \frac{1}{\log\alpha} $ is set as trainable parameter $ \mu_{k}^{i} $ to speed up convergence.
    
	\begin{subequations}
		\begin{align}
		&{\mathbf{W}}_{BB,k}^{i} = \mathbf{T}^{i}_{w,k} \Big((\sum\limits_{v=1}\limits^{K}\mathbf{H}_{eq,k}\mathbf{F}_{BB,v}^{i-1}\mathbf{F}_{BB,v}^{i-1H}\mathbf{H}_{eq,k}^{H} + \sigma_{k}^2\mathbf{I})^{-1} \notag \\
		&\quad\quad\quad\quad\quad \!\mathbf{H}_{eq,k}\mathbf{F}_{BB,k}^{i-1}\Big)
		+ \mathbf{Q}_{w,k}^{i},\label{W^i,k}\\
		&{\mathbf{\lambda}}_{k,l}^{i} = \mathbf{T}^{i}_{\lambda,k}
		(\mu_{k}^{i}\alpha^{\mathbf{t}_{k,l}^{i}}) + \mathbf{q}_{\lambda,k}^{i},\label{l^i_k}\\
		&\mathbf{f}_{BB,k,l}^{i} \!\!=\!\!\mathbf{T}_{f,k}^{i}\!\Big(\lambda_{k,l}^{i}((\!\mathbf{C}^{i}_{k})^{+} \mathbf{B}_{f,k}^{i} \!\!+\!\! (\mathbf{D}_{f,k}^{i})^{-}\!) \mathbf{H}_{eq,k}^{H}\! \mathbf{w}_{BB,k,l}^{i}\!\Big) \!\!+ \!\!\mathbf{q}_{f,k}^{i}, \\
		&{\mathbf{t}}_{k,l}^{i} = \mathbf{T}_{t,k}^{i} (\mathbf{t}_{k,l}^{i-1} + \mu_{k}^{i}(1 - \epsilon_{k,l}\alpha^{\mathbf{t}_{k,l}^{i-1}})) + \mathbf{q}_{t,k}^{i}. \label{t^i_k}
		\end{align}
	\end{subequations}

	 Thus, the trainable parameters of the digital NN are $ \mathbf{\Omega} \triangleq \bigcup_{i=1}^{L_{h}} \{\mathbf{B}_{f,k}^{i}, \mathbf{D}_{f,k}^{i}\} \cup \bm{\varpi}^{i}_{B} \cup \mu_{k}^{i} $, where $ L_{h} $ is the number of layers of the digital NN.
	 All the trainable parameters of the HBDUN are denoted as $ \mathbf{\Upsilon}_{B} \triangleq \mathbf{\Psi} \cup \mathbf{\Omega} $. In addition, to avoid gradient explosion and satisfy the power constraint, we normalize $ \mathbf{F}_{BB,k}^{i} $ by $ N_{s} $ at each layer to $ \frac{\sqrt{N_{s}}}{\parallel\mathbf{F}_{RF,k}\mathbf{F}_{BB,k}^{i}\parallel_{F}} \mathbf{F}_{BB,k}^{i}$. 

	It is interesting to investigate the relationship of the analog beamformers in the proposed deep-unfolding algorithm and the conventional SSCA-based algorithm. In SSCA, $ \bm{\phi} $ is updated based on the gradient $ \dfrac{\partial r_{0}((\bm{\phi}, \bm{M});\mathbf{H})}{\partial \bm{\phi}}|_{\bm{\phi} = \bm{\phi}^{i}} $ while $ \mathbf{\Psi} $ is updated in the HBDUN based on the gradient
 	\begin{equation}
 	\begin{split}
 	\dfrac{\partial \mathcal{L}_{1}(\mathbf{\Upsilon}_{B};\mathbf{H})}{\partial \mathbf{\Psi}}|_{\mathbf{\Psi} = \mathbf{\Psi}^{i}} = &\dfrac{\partial r_{0}((\mathbf{\Psi}, \mathcal{P}_{1}(\{\mathbf{\Psi}, \mathbf{\Omega}\};\mathbf{H}));\mathbf{H})}{\partial \mathbf{\Psi}}|_{\mathbf{\Psi} = \mathbf{\Psi}^{i}}
 	\\
 	=&\dfrac{\partial r_{0}((\mathbf{\Psi}, \mathcal{P}_{1}(\{\mathbf{\Psi}^{i}, \mathbf{\Omega}\};\mathbf{H}));\mathbf{H})}{\partial \mathbf{\Psi}}|_{\mathbf{\Psi} = \mathbf{\Psi}^{i}}\\
 	+&{(\dfrac{\partial r_{0}}{\partial \mathcal{P}_{1}})}^{T}\dfrac{\partial \mathcal{P}_{1}(\{\mathbf{\Psi}^{i}, \mathbf{\Omega}\};\mathbf{H})}{\partial \mathbf{\Psi}}|_{\mathbf{\Psi} = \mathbf{\Psi}^{i}}.
 	\end{split}    	\label{gradient}
 	\end{equation}
 	The gradient of the SSCA algorithm is the same as the first term in (\ref{gradient}) except that the hybrid beamformers are acquired by the HBDUN. The second term is the gradient of the NN which only exists in the deep-unfolding NN but not in the SSCA algorithm. This term further ties the analog and digital beamformers.

  	{\section{Extensions of the Mixed-Timescale Deep-Unfolding Framework}\label{extension}
	\begin{figure}[t]
  		\begin{centering}
  			\includegraphics[width=0.5\textwidth]{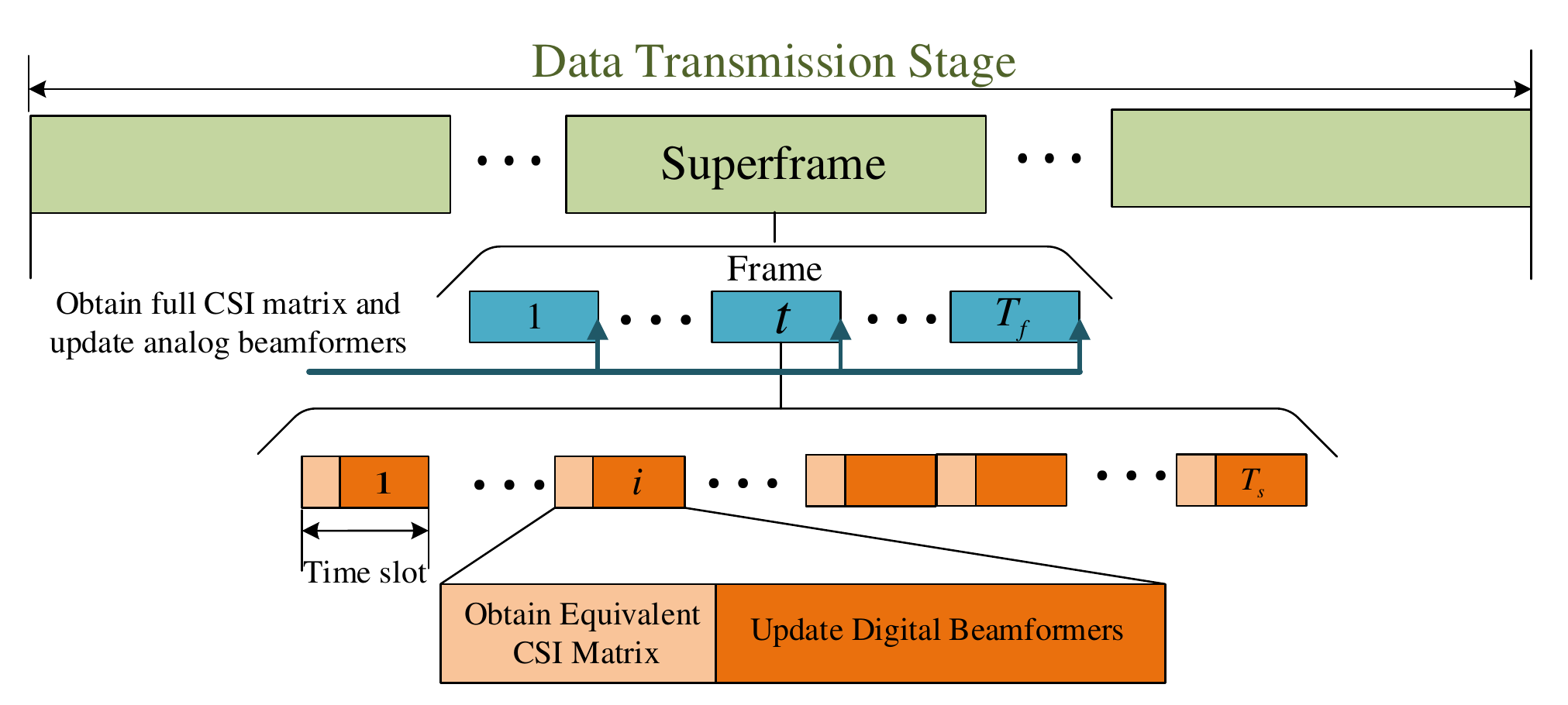}
  			\par\end{centering}
  		\caption{Frame structure of the online mixed-timescale scheme.}
  		\label{twotime_online}
  	\end{figure}
  
	\begin{figure*}[t]
		\begin{centering}
			\includegraphics[width=0.8\textwidth]{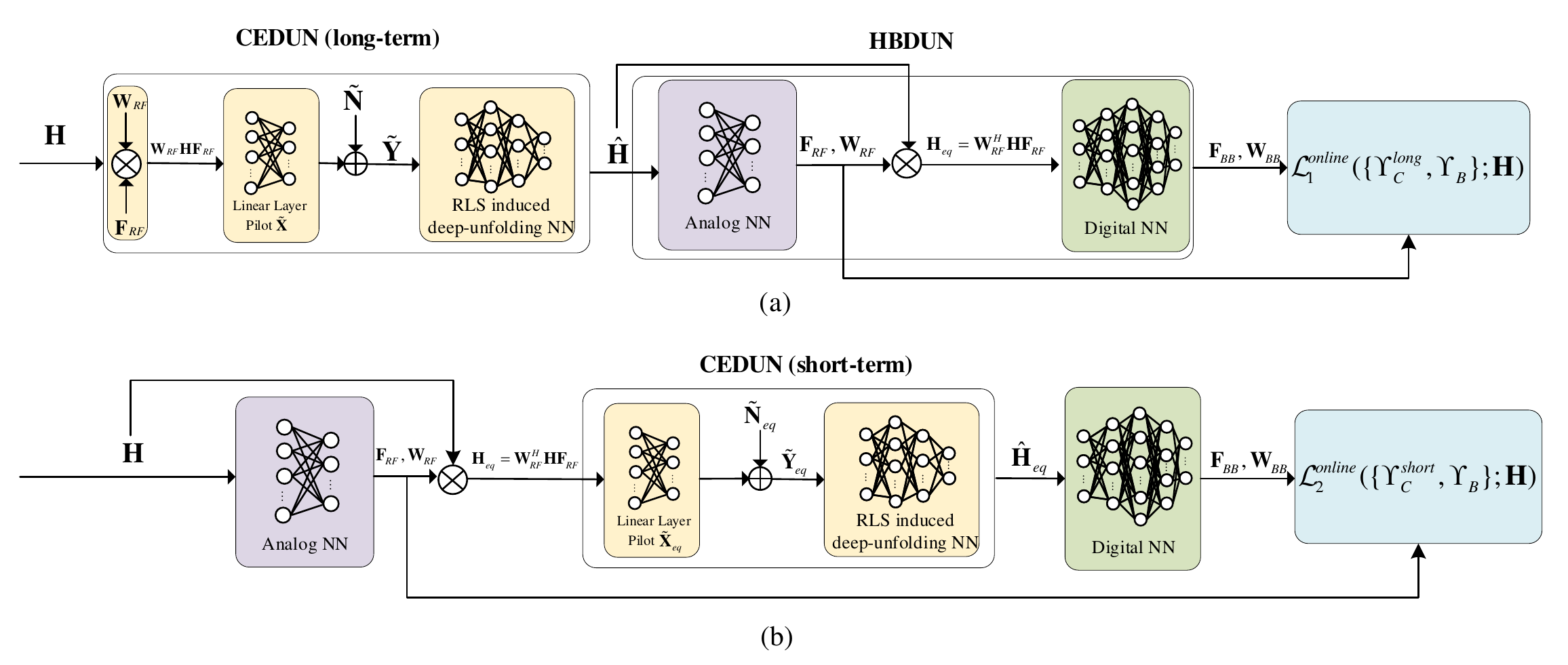}
			\par\end{centering}
		\caption{Structure of the proposed online deep-unfolding framework: (a) Long-term deep-unfolding NN; (b) Short-term deep-unfolding NN.}
		\label{online_joint_NN}
	\end{figure*}
  	In this section, we extend the framework proposed in Section \ref{framework} for other scenarios. In particular, we propose a mixed-timescale deep-unfolding framework where the analog
  	beamformers are optimized online. Besides, we propose an end-to-end deep-learning based framework and employ it in the FDD system, where channel quantization and feedback are considered in the design.
  	\subsection{Online Mixed-Timescale Deep-Unfolding Framework}
  	\subsubsection{Scenarios for Offline and Online Optimization}
  	It is appropriate for optimizing analog beamformers offline in scenarios where CSI characteristics are known and a number of channel samples can be collected before data transmission stage. Besides, when channel statistics change, we need to collect channel samples to fine-tune the deep-unfolding NN. During the time it takes to collect channel samples, the analog beamformers are not optimized. Thus, if it is difficult to collect channel samples, the offline optimization is not appropriate.
  	If we do not know the CSI statistics or it is difficult to collect a large number of channel samples, the analog beamformers need to be updated online. In this way, we collect channel samples and optimize analog beamformers at the same time during data transmission. The performance gradually improves as the number of collected samples increases and eventually converges when there are enough channel samples.
   	\subsubsection{Online Mixed-Timescale Scheme}
	In the online mixed-timescale scheme, we optimize long-term analog beamformers and short-term digital beamformers at different timescales. The frame structure is shown in Fig. \ref{twotime_online}. In particular, at the end of each frame, we obtain full CSI samples and optimize analog beamformers using the long-term deep-unfolding NN while at each time slot, we obtain equivalent CSI matrices to optimize the digital beamformers using the short-term deep-unfolding NN.
  	\subsubsection{Structure of the Online Deep-Unfolding Framework}
	 We propose a mixed-timescale deep-unfolding framework for online training, the structure of which is shown in Fig. \ref{online_joint_NN}. At the end of each frame, we train the long-term deep-unfolding NN, as shown in Fig. \ref{online_joint_NN}(a), to optimize the analog beamformers. The long-term deep-unfolding NN consists of long-term CEDUN for full CSI estimaiton and HBDUN for hybrid beamforming. At each time slot, we train the short-term deep-unfolding NN shown in Fig. \ref{online_joint_NN}(b) to optimize the digital beamformers. The training process is similar to that in Section \ref{twostage}.
  	\begin{figure*}[t]
	 	\begin{centering}
	 		\includegraphics[width=0.85\textwidth]{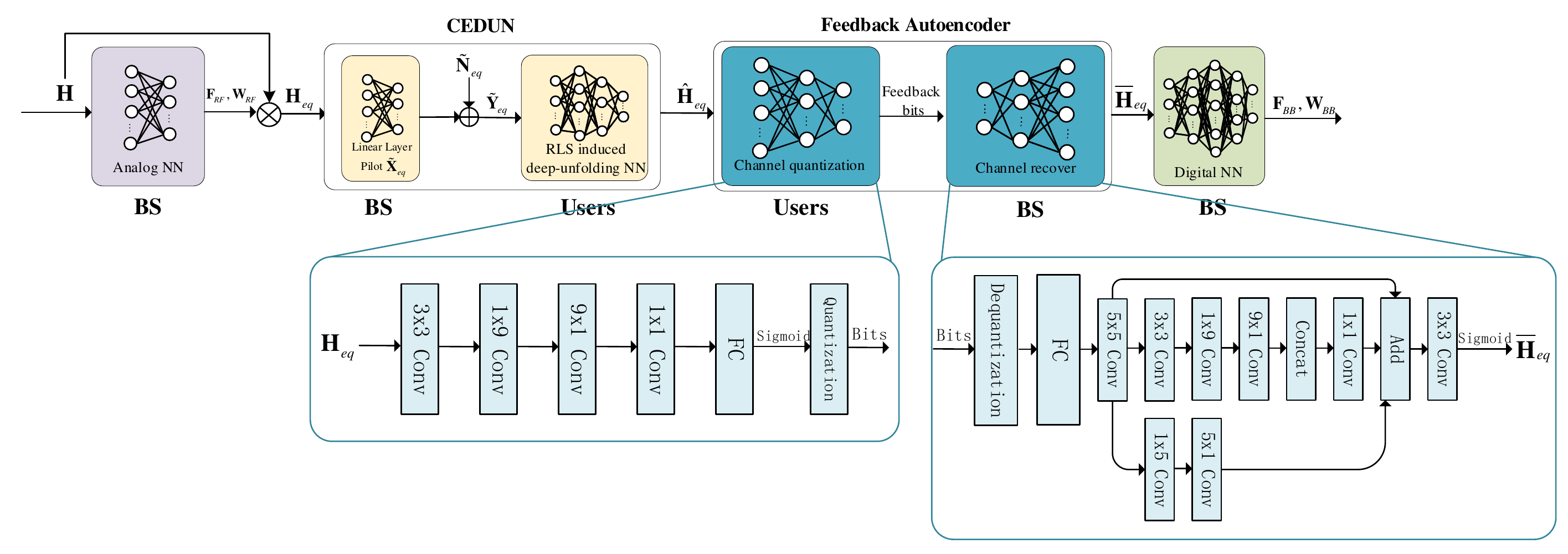}
	 		\par\end{centering}
	 	\caption{Architecture of the deep-unfolding framework for FDD systems.}
	 	\label{FDD_joint_NN}
	 \end{figure*}
	\subsection{End-to-End Deep-Learning Framework for FDD systems}
	\subsubsection{Channel Estimation and Feedback for FDD Systems}
	In FDD systems, there is no channel reciprocity. To acquire downlink CSI, the BS needs to send pilots to the users. Then the users estimate the CSI matrix $ \mathbf{H} $ which is quantized as bits and fed back to the BS. The BS recovers the CSI matrix for hybrid beamforming based on the feedback bits. During the data transmission in the mixed-timescale scheme, we fix the analog beamformers and only estimate the equivalent CSI at each time slot. Thus, we only need to feed the equivalent CSI back to the BS.
	\subsubsection{Structure of the Deep-Learning Based Framework for FDD systems}
	 In an end-to-end FDD system, the modules of channel estimation and hybrid beamforming are required to have strong generalization and interpretability, which can be designed by deep-unfolding approaches. However, the module of channel feedback, which is used to extract and compress channel features and recover the channel, does not need to have strong interpretability and generalization for signal-to-noise ratio (SNR). Black-box NNs can effectively extract channel features and reduce feedback overhead \cite{feedback0}, which are suitable for channel feedback design. Besides, it is difficult to unfold the conventional algorithm \cite{feedback1,feedback2,feedback3} for channel quantization and feedback due to the non-differentiable procedures. Thus, we propose an autoencoder based on the CRNet in \cite{feedback4} which significantly compresses the channel matrix and reduces the transmission overhead. The structure of the feedback autoencoder is shown in Fig. \ref{FDD_joint_NN}, and consists of several convolution layers and fully connected (FC) layers. The dimension of the equivalent CSI matrix is much lower than that of full CSI matrix and less information needs to be fed back. 
	
	We jointly design the feedback autoencoder and deep-unfolding NNs in the proposed deep-learning framework. Here the RLS induced deep-unfolding NN and the channel quantization module are deployed at the users while the pilot training NN, channel recovery module, and HBDUN are deployed at the BS.
	
	The performance of the autoencoder only depends on the equivalent CSI statistics but not on the transmit power or the channel noise. Thus, the autoencoder is not required to have a strong generalization ability for SNR and can be trained separately. The training process is thus as follows. 
	\begin{itemize}
		\item \textbf{Training Stage I:} First we jointly train the deep-unfolding NN except for the autoencoder according to Section III.A. 
		\item \textbf{Training Stage II:} Then we separately train the autoencoder for channel quantization and recovery. We fix the analog beamformers and obtain the equivalent CSI matrix $ \mathbf{H}_{eq} $, which is the input of the autoencoder. Its output is the recovered CSI matrix $ \bar{\mathbf{H}}_{eq} $. The loss function is normalized mean square error (NMSE), which is defined as follows:
		\begin{equation}
		\text{NMSE} = \dfrac{{\parallel \bar{\mathbf{H}}_{eq}-{\mathbf{H}}_{eq} \parallel}^2_2}{\parallel{\mathbf{H}}_{eq}\parallel^2_2}.
		\end{equation}
		\item \textbf{Training Stage III:} Finally, we load the well-trained models into the deep-unfolding NN and the autoencoder and cascade them as shown in Fig. \ref{FDD_joint_NN}. Then we jointly train them with the sum rate as the loss function.
	\end{itemize}
	}
 	\section{Computational Complexity and Performance Analysis}
 	\label{analysis}
 	In this section, we first develop a black-box NN for the JCEHB  design for comparison. Then we analyze the computational complexity of the conventional algorithms, the proposed deep-unfolding NN and the black-box NN. We further analyze the convergence of the proposed deep-unfolding NN based algorithm.
 	\subsection{The Black-Box NNs for JCEHB Design}
\begin{figure}[t]
 		\begin{centering}
 			\includegraphics[width=0.46\textwidth]{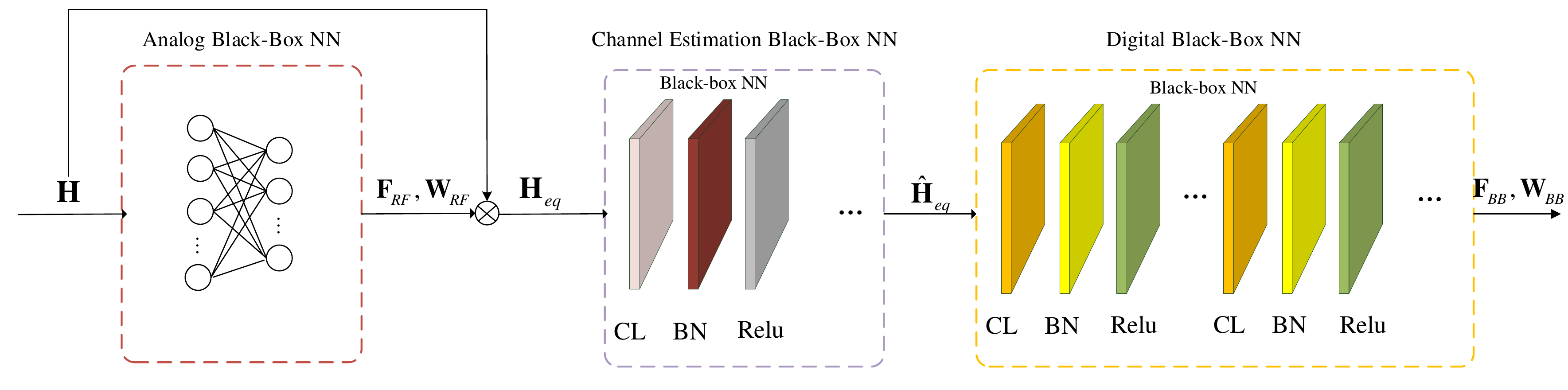}
 			\par\end{centering}
 		\caption{Architecture of the black-box NN.}
 		\label{blackbox}
 	\end{figure}

 	 The structure of the black-box NN is shown in Fig. \ref{blackbox}. For the analog beamforming design, we propose the analog black-box NN which sets the phase of analog beamformers as trainable parameters similar to the proposed deep-unfolding NN. For channel estimation and digital beamforming design, we propose the channel estimation and digital black-box NN, which adopt the DNN that consists of conventional FC layers and batch normalization (BN) layers. ReLU is used as the activation function. Specifically, the full CSI sample $ \mathbf{H}_{k} $ passes through the analog black-box NN and analog beamformers $\{ \mathbf{F}_{RF,k}, \mathbf{W}_{RF,k}\} $ are obtained. Then we obtain the equivalent CSI matrix $ \mathbf{H}_{eq,k} $ which passes through the channel estimation black-box NN that outputs the estimated equivalent channel matrix $ \hat{\mathbf{H}}_{eq,k} $. Finally, $ \hat{\mathbf{H}}_{eq,k} $ passes through the digital black-box NN which outputs the digital beamformers $\{ \mathbf{F}_{BB,k}, \mathbf{W}_{BB,k}\} $. The input of the black-box NN is the full CSI matrix and the outputs are hybrid beamformers. The loss function of the black-box NN is the system sum rate.
 	{
 	\subsection{Computational Complexity}
 	We analyze the computational complexity of the conventional RLS and SSCA algorithm, the proposed deep-unfolding NN, and the black-box NN.
 	\subsubsection{Conventional Algorithms}
 	The computational complexity of the RLS algorithm is
 	\begin{equation}
    \mathcal{O}(L_{r}K({N_{t}^{RF}})^2) ,
 	\end{equation} 
 	 where $ L_{r} $ is the number of iterations. The computational complexity of the SSCA algorithm is
 	\begin{equation}
 		  \mathcal{O}(L_{s}\!(KN_{s} (\!{N_{t}^{RF}}\!)^{3} \!+\! K^{2\!}N_{s}^{2}(\!{N_{t}^{RF}}\!)^{2}N_{r}^{RF} \!+\! K\!N_{s}(\!{N_{t}^{RF}}\!)^{2}\!N_{r}^{RF}\!)),
 	\end{equation}
 	where $ L_{s} $ is the number of iterations.
 	\subsubsection{Proposed Deep-Unfolding NN}
 	The computational complexity of the CEDUN is 
 	\begin{equation}
    \mathcal{O}(L_{c}K({N_{t}^{RF}})^2),
 	\end{equation}  
 	where $ L_{c} $ $ (L_{c}\ll L_{r}) $ represents the number of layers of the CEDUN. The computational complexity of the HBDUN is
 	\begin{equation}
	 	  \mathcal{O}\!(L_{h}\!(K\!N_{s} (\!{N_{t}^{RF}}\!)^{2.37} \!\!+\! K^{2}\!N_{s}^{2}(\!{N_{t}^{RF}}\!)^{2}\!N_{r}^{RF}\! \!+\! K\!N_{s}(\!{N_{t}^{RF}}\!)^{2}\!N_{r}^{RF})\!),
 	\end{equation}
 	where $ L_{h} $ $ (L_{h}\ll L_{a}) $ represents the number of layers of the HBDUN. Compared to the traditional algorithms, the proposed deep-unfolding NN has a lower computational complexity:
 	\begin{itemize}
 		\item The number of layers of the proposed deep-unfolding NN is reduced, i.e., $ L_{c}\ll L_{r}, L_{h}\ll L_{s} $.
 		\item The HBDUN approximates the matrix inversion operation with computational complexity $ \mathcal{O}(n^{3}) $ by matrix multiplication operation with lower computational complexity $ \mathcal{O}(n^{2.37}) $.
 	\end{itemize}
    \subsubsection{Black-Box NN}
 	The computational complexity of the proposed black-box NN is
 	\begin{equation}
 		 \mathcal{O}(\sum_{l=1}^{L_{b}-1}K F_{c,l}F_{c,l+1} + \sum_{l=1}^{L_{d}-1}KF_{h,l}F_{h,l+1}) , 
 	\end{equation}
 	where $ L_{b} $ and $ L_{d} $ denote the numbers for fully connected layers of channel estimation and digital black-box NN, respectively, and $ F_{c,l} $ and $ F_{h,l} $ denote the output sizes of the $ l $-th layer for channel estimation and hybrid beamforming black-box NN, respectively.

 	\subsection{Performance Analysis for the Deep-unfolding NN}
 	We show that the performance of one layer of the deep-unfolding NN can approach that of several iterations of the conventional iterative algorithm. We take the HBDUN as an example; the CEDUN can be analyzed in the same way. 
 	
 	We consider the update of $ \mathbf{t} $ as an example, where $ \mathbf{W}_{BB} $ and $ \mathbf{F}_{BB} $ are treated as constants. As shown in Algorithm 2, in the $ i $-th iteration of the SCA algorithm, we have the following mapping from $ \mathbf{t}_{k,l}^{i} $ to $ \mathbf{t}_{k,l}^{i+2} $:
 	\begin{equation}
	\mathbf{t}_{k,l}^{i+2} \!=\! \mathbf{t}_{k,l}^{i} \!+\! \dfrac{2}{\log\alpha} \!- \dfrac{1}{\log\alpha}(\!1 \!+\! \alpha^{\frac{1}{\log\alpha}(1-\varepsilon_{k,l}^{i+1}\alpha^{\mathbf{t}_{k,l}^{i}})})\varepsilon_{k,l}^{i}\alpha^{\mathbf{t}^{i}_{k,l}}. \label{i+2}
  	\end{equation}
  	Similarly, we have the following mapping from $ \mathbf{t}_{k,l}^{j} $ to $ \mathbf{t}_{k,l}^{j+1} $ in the HBDUN:
  	\begin{equation}
  	\mathbf{t}_{k,l}^{j+1} = \mathbf{T}_{t,k}^{j}\mathbf{t}_{k,l}^{j} - \mathbf{T}_{t,k}^{j}\mu_{k}^{j}\varepsilon_{k,l}^{j}\alpha^{\mathbf{t}^{j}_{k,l}} + \mathbf{q}_{t,k}^{j}. \label{l+1}
  	\end{equation}
  	We need to prove that $ \mathbf{t}_{k,l}^{j+1} $ approaches $ \mathbf{t}_{k,l}^{i+2} $, i.e., $ \| \mathbf{t}_{k,l}^{j+1} -  \mathbf{t}_{k,l}^{i+2}\|^{2} < \zeta$, for any $ \zeta> 0 $. 
  	
  	For deterministic channels, we can demonstrate that there exist trainable parameters $ \mathbf{T}_{t,k}^{j}, \mu_{k}^{j} $ and $ \mathbf{q}_{t,k}^{j} $ that satisfy  $ \| \mathbf{t}_{k,l}^{j+1} -  \mathbf{t}_{k,l}^{i+2}\|^{2} < \zeta$. Based on (\ref{i+2}) and (\ref{l+1}), we obtain the following formulation:
  	\begin{subequations}
  		    	\makeatletter
  		\def\@eqnnum{{\normalfont (\theequation)}} \makeatother
  		\begin{eqnarray}
  		&\mathbf{T}_{t,k}^{j} = 1, \forall k, \\
  		&\mathbf{q}_{t,k}^{j} = \dfrac{2}{\log\alpha},\forall k,\\
  		&\mu_{k}^{j} = \dfrac{1}{\log\alpha}(1 + \alpha^{\frac{1}{\log\alpha}(1-\varepsilon_{k,l}^{i+1}\alpha^{\mathbf{t}_{k,l}^{i}})}),\forall k,
  		\end{eqnarray}
  	\end{subequations}
  	where $ \varepsilon_{k,l}^{i+1} $ can be obtained by $ \mathbf{W}_{BB} $ and $ \mathbf{F}_{BB} $ in the $ i $-th iteration according to (\ref{mse}).
  	For channels that follow a certain distribution, we need to show that $\mathbb{E}_{\mathbf{H}}\{\| \mathbf{t}_{k,l}^{j+1} -  \mathbf{t}_{k,l}^{i+2}\|^{2} \}< \delta . $
  	 By taking $ \mathbf{T}_{t,k}^{j} = 1$ and $ \mathbf{q}_{t,k}^{j} = \dfrac{2}{\log\alpha},  \forall k$, we need to prove that
  	 \begin{equation}
  	  \mathbb{E}_{\mathbf{H}}\{\|(\mu_{k}^{i}  \!-\!  \dfrac{1}{\log\alpha}(1 \!+\! \alpha^{\frac{1}{\log\alpha}(1-\varepsilon_{k,l}^{i+1}\alpha^{\mathbf{t}_{k,l}^{i}})}))\varepsilon_{k,l}^{i}\alpha^{\mathbf{t}^{i}_{k,l}} \|\} \!<\! \zeta. \label{eta}
  	 \end{equation}
  	 Based on \textit{Cauchy–-Schwarz Inequality}, we have 
  	 \begin{equation}
  	 \begin{aligned}
 	 &&\mathbb{E}_{\mathbf{H}}\{(\mu_{k}^{i}  -  \dfrac{1}{\log\alpha}(1 + \alpha^{\frac{1}{\log\alpha}(1-\varepsilon_{k,l}^{i+1}\alpha^{\mathbf{t}_{k,l}^{i}})}))\varepsilon_{k,l}^{i}\alpha^{\mathbf{t}^{i}_{k,l}} \}\\
 	 &&\leq \mathbb{E}_{\mathbf{H}}\{\|(\mu_{k}^{i}  -  \dfrac{1}{\log\alpha}(1 + \alpha^{\frac{1}{\log\alpha}(1-\varepsilon_{k,l}^{i+1}\alpha^{\mathbf{t}_{k,l}^{i}})}))\|\}. \label{equality}
  	 \end{aligned}
  	 \end{equation}
  	Then we set
  	 \begin{equation}
  	 	\mu_{k}^{i} = \mathbb{E}_{\mathbf{H}}\{\dfrac{1}{\log\alpha}(1 + \alpha^{\frac{1}{\log\alpha}(1-\varepsilon_{k,l}^{i+1}\alpha^{\mathbf{t}_{k,l}^{i}})})\}.
  	 \end{equation}
  	 According to \textit{The Law of Large Numbers}, (\ref{equality}) converges to $ 0 $ with a sufficiently large number of channel samples.
  	 Thus there exists the trainable parameter $ \mu_{k}^{j} $ which satisfies (\ref{eta}) for any $ \zeta > 0 $, i.e., the performance of one layer of the proposed deep-unfolding NN can approach that of two iterations of the corresponding conventional iterative algorithm. Indeed, we can extend that the performance generated by one layer of the deep-unfolding NN with a more complex structure can approach that of multiple iterations of the conventional iterative algorithm. 
  	 }
	\section{ Simulation Results}
	\label{Section6:simulations}
	In this section, we verify the effectiveness of the proposed deep-unfolding algorithm based on simulation results.
	\subsection{Simulation Setup}
	The simulation setting is given as follows. We set $ N_{t}=64 $, $ N_{t}^{RF}=16$, $ N_{r}=32 $, $ N _{r}^{RF}=4 $, $ K=4 $, and $ N_{s} = 4 $. We set the SNR as 10 dB and the number of the layers for CEDUN and HBDUN are $ 16 $ and $ 5 $, respectively. 	
	The CSI matrix is given by
	\begin{equation} 
	\mathbf{H} = \sqrt{\dfrac{N_{t}N_{r}}{N_{c}N_{ray}}}\sum\limits_{i=1}^{N_{C}}\sum\limits_{l=1}^{N_{Ray}} \alpha_{il} \mathbf{a}_{r}(\phi_{il}^{r}) \mathbf{a}_{t}^{H}(\phi_{il}^{t}),\label{channel}
	\end{equation}
	where $ N_c $ and $ N_{ray} $ are the number of clusters and propagation rays, respectively, $\alpha_{il}\sim \mathcal{CN}(0,\sigma_{\alpha}^{2})$ is the complex gain, and $\phi_{il}^{r}$ and $\phi_{il}^{t}$ denote the angle of arrival (AoA) and angle of departure (AoD) for the $l$-th ray in the $i$-th cluster, respectively. The $\mathbf{a}_{r}(\phi_{il}^{r})$ and $\mathbf{a}_{t}(\phi_{il}^{t})$ denote the receive and transmit array response vectors, respectively, and for a uniform
	linear array with $N$ antenna elements and angle $ \phi $, the response vector can be expressed as
	\begin{equation}
	\mathbf{a}(\phi)=\frac{1}{\sqrt{N}}\big[1, e^{-j 2\pi \frac{d}{\lambda} \sin(\phi)}, \cdots , e^{-j 2\pi \frac{d}{\lambda}(N-1) \sin(\phi)}  \big]^{T}, \label{angle}
	\end{equation}
	\hspace{-1mm}where $d$ and $\lambda$ denote the distances between the adjacent antennas and carrier wavelength, respectively. We set $ N_c = 4, N_{ray} = 2, \sigma_{\alpha}^{2} = 0.1, \phi_{il}^{r} \sim \mathcal{U}(-\pi/3, \pi/3) $ and $ \phi_{il}^{t} \sim \mathcal{U}(-\pi/3, \pi/3) $. {We use Python (version 3.6) as the programming language and use the library of Pytorch (version 1.8.1) to build the deep learning framework. Besides, the simulations are carried out on a computer with Intel i5 CPU running at 2.8GHz and with 16GB RAM.}

	We provide the following benchmarks as comparison:
	\begin{itemize}
		\item  Joint design NN: The proposed mixed-timescale deep-unfolding NN which jointly trains the CEDUN and HBDUN.
		\item Separate Design NN: The proposed mixed-timescale deep-unfolding NN which separately trains the CEDUN and HBDUN with the minimization of MSE and the maximization of sum rate as the loss function, respectively.
		\item HBDUN: The proposed deep-unfolding NN for the design of hybrid beamforming with perfect CSI.
		\item RLS-SSCA: The cascaded RLS algorithm for the design of channel estimation and the SSCA algorithm for the hybrid beamforming design.
		\item RLS-ZF: The cascaded RLS algorithm for the design of channel estimation and the zero-forcing (ZF) algorithm for the hybrid beamforming design.
		\item Black-box NN: The proposed black-box NN for the JCEHB design.
	\end{itemize}

    \subsection{Pilot Overhead}
	\begin{figure}[t]
		\begin{centering}
			\includegraphics[width=0.4\textwidth]{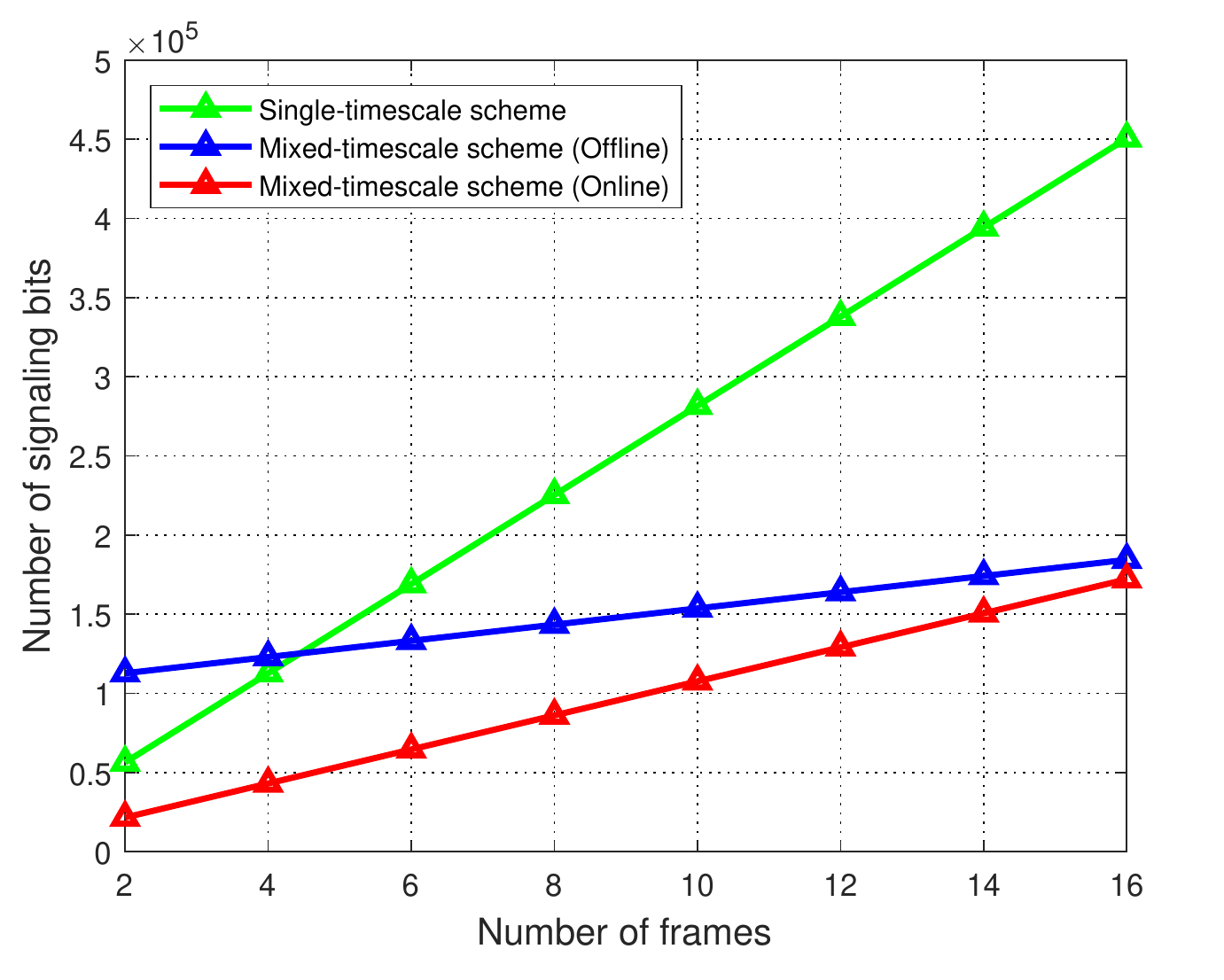}
			\par\end{centering}
		\caption{Signaling bits versus the number of frames.}
		\label{overhead}
	\end{figure}
	
	We compare the signaling bits for the feedback of pilots versus the number of frames for the single-timescale and mixed-timescale schemes in Fig. \ref{overhead}. For the full CSI estimation, the users need to transmit the pilots $ \tilde{\mathbf{W}}_{RF,k}\in \mathbb{C}^{N_{r}\times N_{r}^{RF}} $, $ \tilde{\mathbf{F}}_{RF,k}\in \mathbb{C}^{N_{t}\times N_{t}^{RF}} $, and $ \tilde{\mathbf{X}}_{k}\in \mathbb{C}^{N_{t}^{RF} \times L} $. The number of feedback signaling bits for pilots is given by 	
	\begin{equation}
	Q_{f} = q(N_{r}N_{r}^{RF}+N_{t}N_{t}^{RF}+N_{t}^{RF}L),
	\end{equation}
	where $ q $ denotes the number of bits for quantifying each element of pilot matrix. For the equivalent CSI estimation, the users only need to feed back the pilot $ \tilde{\mathbf{X}}_{eq,k}\in \mathbb{C}^{N_{t}^{RF} \times L} $ and the number of feedback signaling bits is given by
	\begin{equation}
	Q_{eq} = qN_{t}^{RF}L.
	\end{equation}
	For the single-timescale scheme, we need to estimate full CSI at each time slot. Thus, the number of feedback signaling bits for the single-timescale scheme over each superframe is given by 	
	\begin{equation}
	Q_{single} = T_fT_sQ_{f},
	\end{equation}
	where $ T_f $ is the number of frames in a superframe and  $ T_s $ is the number of time slots in each frame.
	
 In the offline mixed-timescale scheme, we need to estimate CSI statistics for offline training. We can estimate full CSI and employ the expectation-maximum (EM) algorithm \cite{EM} or maximum likelihood (ML) algorithm to estimate the distribution function of channel parameters, e.g., AoAs, AoDs, and complex gains. During the data transmission stage, we only need to estimate the equivalent CSI at each time slot. Thus, the pilot overhead of the offline mixed-timescale scheme over each superframe is given by
	\begin{equation}
	Q_{offline} = T_fT_sQ_{eq} + N_{sample}Q_f,
	\end{equation}
	where $ N_{sample} $ represents the number of estimated channel samples for CSI statistics estimation.
	
	In the online mixed-timescale scheme, we need to estimate full CSI at the end of each frame and estimate equivalent CSI at each time slot. Thus, the pilot overhead of the online mixed-timescale scheme over each superframe is given by
	\begin{equation}
	Q_{online} = T_f(T_sQ_{eq} + Q_f).
	\end{equation}
	
	In Fig. \ref{overhead}, we consider a superframe where channel statistics do not change. The number of frames in a superframe depends on the coherence time of the channel. When the CSI statistics change fast, the number of frames in a superframe is small and the CSI statistics need to be estimated frequently in the offline scheme, which leads to high pilot overhead. In contrast, when the CSI statistics change slow, the number of frames is large and the pilot overhead can be reduced in the offline scheme. 
  	Here, we assume that the superframe consists of $ 16 $ frames, i.e., $ T_f = 16 $.
  	
	It can be observed that when the number of frames is samll, e.g., the number of frames is 2, the pilot overhead of offline mixed-timescale scheme is larger than that of the other two schemes due to the estimation for CSI statistics. 
  	As the number of frames increases, the pilot overhead of three schemes increase. We can observe that when the number of frames is large, the pilot overhead of the single-timescale scheme is much larger than that of the mixed-timescale scheme since we need to estimate the full CSI at each time slot in the single-timescale scheme. In addition, the pilot overhead of online and offline mixed-timescale is almost the same.	
  	Thus, taking both of the offline training stage and the data transmission stage into consideration, the pilot overhead of the mixed-timescale schemes is much smaller than that of the conventional single-timescale schemes.

    \begin{figure}[t]
    	\begin{centering}
    		\includegraphics[width=0.43\textwidth]{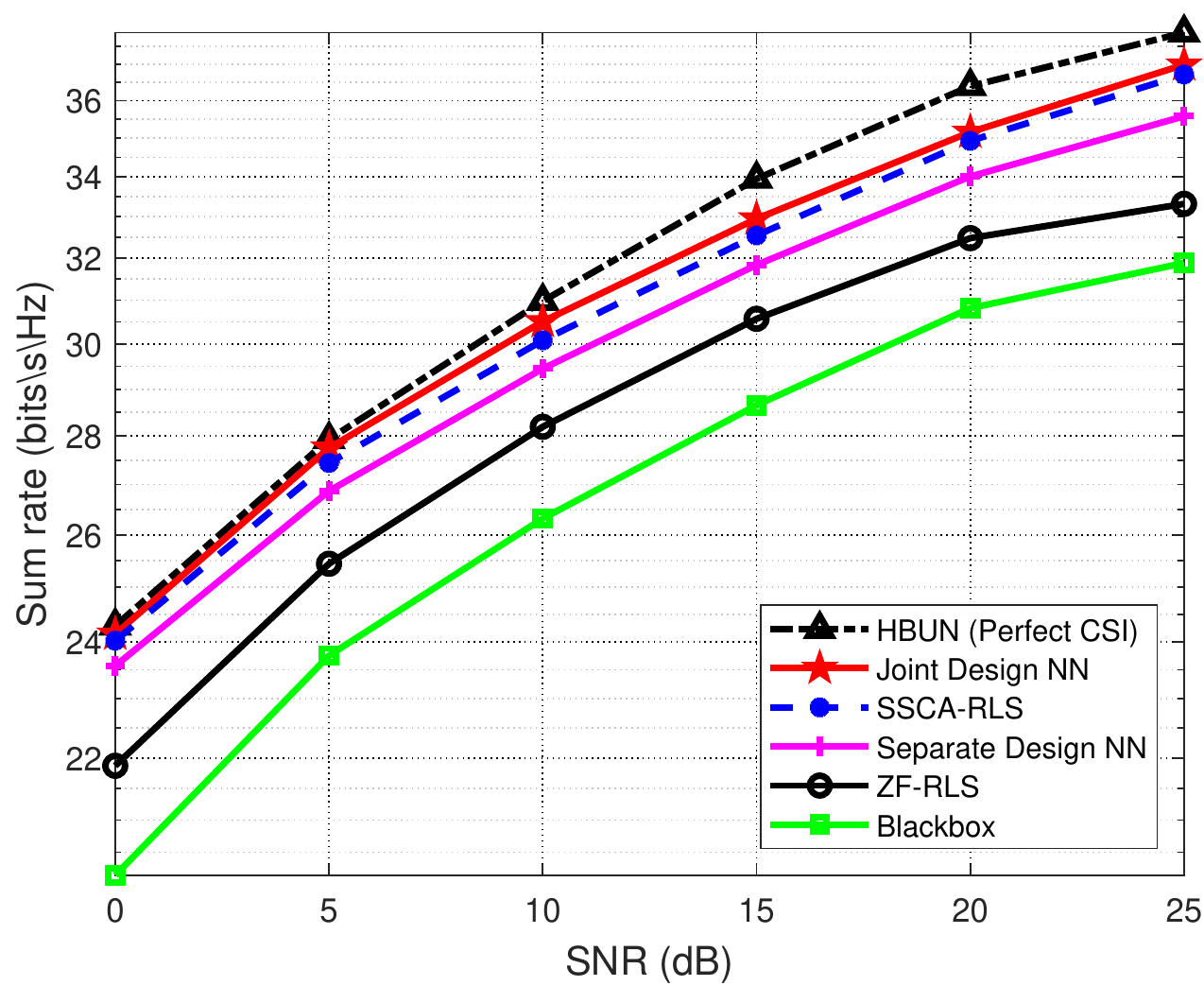}
    		\par\end{centering}
    	\caption{The sum rate versus the SNR.}
    	\label{SNR}
    \end{figure} 
    \begin{figure}[t]
	\begin{centering}
		\includegraphics[width=0.43\textwidth]{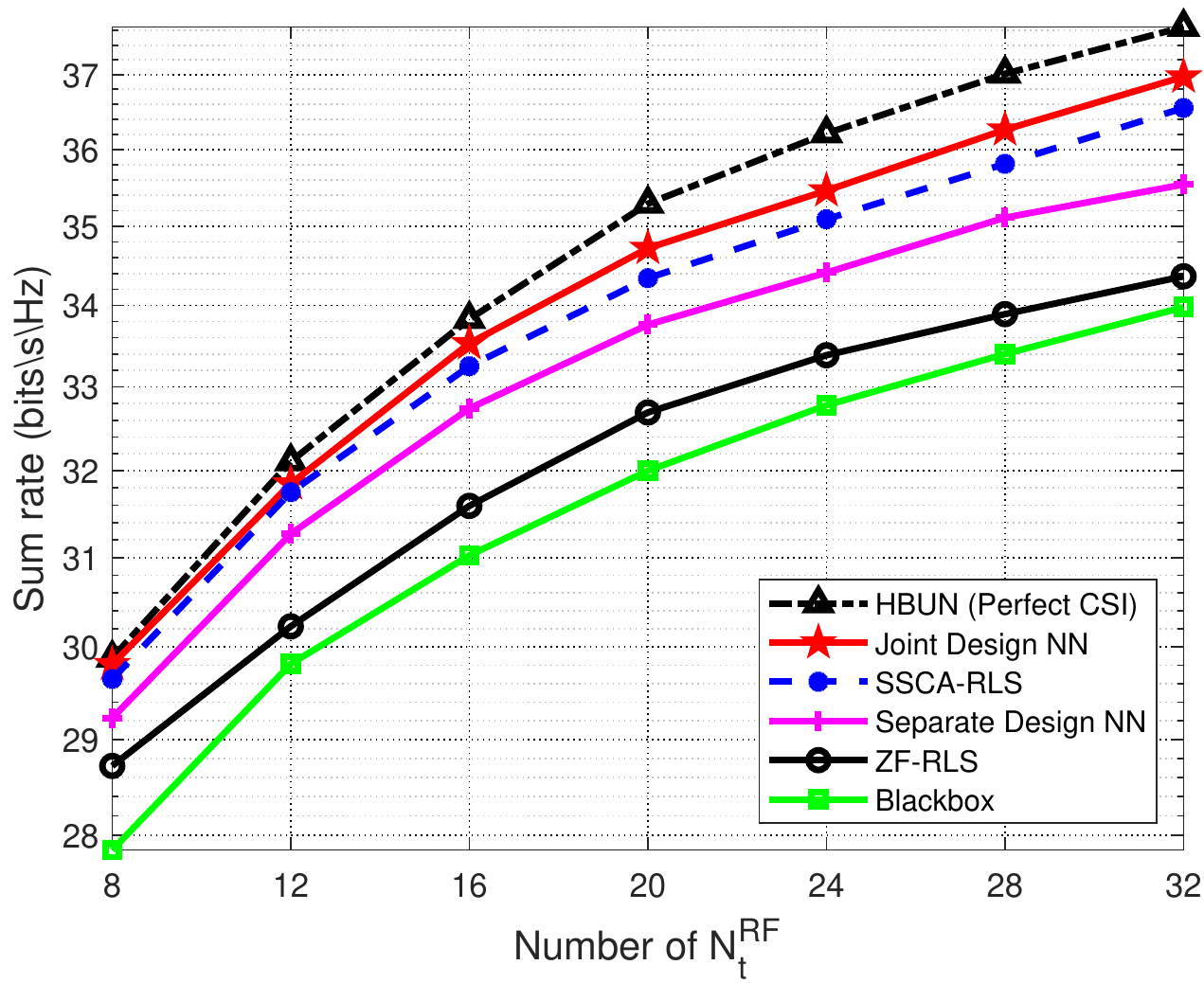}
		\par\end{centering}
	\caption{The sum rate versus the number of RF chains $ N_{t}^{RF} $.}
	\label{NtRF}
	\end{figure}
 	\subsection{System Sum Rate}
	Fig. \ref{SNR} illustrates the sum rate of the proposed network and the benchmarks for different values of SNR. Firstly it can be seen that for all algorithms, the sum rate increases gradually with SNR. The proposed separate deep-unfolding NN can achieve comparable performance compared to the conventional SSCA and RLS algorithms, which indicates the effectiveness of our proposed deep-unfolding NN. The results also illustrate the superiority of the joint design compared to the separate NN. Besides, both of the proposed deep-unfolding NNs significantly outperform the black-box NN. It is mainly because the proposed deep-unfolding NNs are designed based on the iterative optimization algorithms.
 	    
    Fig. \ref{NtRF} indicates the sum rate of the proposed algorithm and the benchmarks for different numbers of the RF chains $ N_{t}^{RF} $. We can see that the sum rate achieved by different algorithms increases with the number of RF chains. Moreover, it is observed that the sum rate of the proposed joint design NN exceeds that of the conventional iterative algorithms.
    Besides, we can see that the performance gap between the joint design NN and separate design NN becomes larger as $ N_{t}^{RF} $ increases, which illustrates the performance gain of the joint design NN.
    
	\begin{table*}[htbp]
		\centering
		\caption{The sum rate versus different numbers of $ K $.}
		\begin{tabular}{c|cccccc}
			\hline
			$K$&1&2&3&4&5&6 \\ 
			\hline
			RLS-SSCA (bits/s/Hz)&$10.18$&$19.55$&$24.78$&$29.64$&$35.84$&$42.63$ \\
			\hline
			Separate Design NN&$98.33\%$&$97.23\%$&$97.06\%$&$96.77\%$&$95.60\%$&$95.33\%$ \\
			\hline
			Joint Design NN&$103.53\%$&$102.32\%$&$100.31\%$&$100.07\%$&$99.49\%$&$98.78\%$ \\
			\hline
			Black-box NN&$86.95\%$&$85.65\%$&$85.01\%$&$83.99\%$&$83.42\%$&$81.94\%$ \\
			\hline 
		\end{tabular}
		\label{userk}
	\end{table*}
	
    Table \ref{userk} manifests the sum rate versus the number of users $ K $. We normalize the sum rate of NNs by the corresponding value of the RLS-SSCA algorithm. We can see that when $ K $ is small, e.g., $ K=2 $, the proposed separate design NN can achieve $ 97.23\% $ performance of the conventional iterative algorithms and the joint design NN outperforms the conventional iterative algorithms. It can be seen that the sum rates of deep-unfolding NNs and black-box NNs both degrade with the increase of $ K $. It is mainly because as $ K $ increases, the problem turns more complex and it is difficult for NNs to find a satisfactory solution.
	\begin{table*}[htbp]
		\centering
		\caption{The sum rate versus the number of training samples.}
		\begin{tabular}{cccccccccc}
			\hline
			Training samples &200&300&400&500&600&700&800&900&1000\\ 
			\hline
			Deep-unfolding NN&$90.14\%$&$94.41\%$&$96.25\%$&$98.09\%$&$99.44\%$&$100.11\%$&$101.30\%$&$101.31\%$&$101.31\%$ \\
			\hline 
			Training samples &2000&3000&4000&5000&6000&7000&8000&9000&10000 \\ 
			\hline
			Black-box NN&$76.77\%$&$77.59\%$&$78.90\%$&$80.52\%$&$82.27\%$&$83.22\%$&$84.62\%$&$85.44\%$&$85.57\%$ \\
			\hline
			\label{sample}
		\end{tabular}
	\end{table*}
	
	Table \ref{sample} presents the sum rate versus the number of training samples. We normalize the results by the sum rate of the RLS-SSCA algorithm. It is obvious that the samples required by the deep-unfolding NN for training are much fewer than that of the black-box NN. In reality, it is challenging to obtain a large quantity of training samples. Thus, the proposed deep-unfolding NN is more practical.
	\begin{table*}[htbp]
		\centering
		\caption{The sum rate versus the number of HBDUN/SSCA layers/iterations.}
		
		\begin{tabular}{cccccccc}
			\hline
			The number of layers of HBDUN&3&4&5&6&7&8&9\\ 
			\hline
			sum rate &$85.14\%$&$92.65\%$&$101.25\%$&$101.79\%$&$101.79\%$&$100.86\%$&$101.02\%$ \\
			\hline 
		    The number of layers of SSCA &30&35&40&45&50&55&60\\ 
		    \hline
		    sum rate &$84.84\%$&$89.44\%$&$92.43\%$&$96.99\%$&$98.86\%$&$100\%$&$100\%$ \\
			\hline 
			\label{layer_sca}
		\end{tabular}
	\end{table*}

	\begin{table*}[htbp]
	\centering
	\caption{The sum rate versus the number of CEDUN/RLS layers/iterations.}
	
	\begin{tabular}{cccccccc}
		\hline
		The number of layers of CEDUN&8&10&12&14&16&18&20\\ 
		\hline
		sum rate &$92.18\%$&$95.44\%$&$98.15\%$&$99.09\%$&$101.25\%$&$101.46\%$&$101.5\%$ \\
		\hline 
		The number of layers of RLS &40&50&60&70&80&90&100\\ 
		\hline
		sum rate &$79.64\%$&$86.69\%$&$91.85\%$&$95.47\%$&$100\%$&$100\%$&$100\%$ \\
		\hline 
		\label{layer_rls}
	\end{tabular}
\end{table*}
	Table \ref{layer_sca} and Table \ref{layer_rls} indicate the sum rate versus the number of layers/iterations for deep-unfolding NN and conventional iterative algorithms. We normalize the results by the sum rate which is achieved by the conventional algorithm with $ 90 $ layers of the RLS algorithm and $ 70 $ layers of the SSCA algorithm. We can see that the deep-unfolding NNs achieve better performance than the conventional algorithms with much less number of layers. It is observed that with more layers, the sum rate achieved by the joint deep-unfolding NN improves firstly and then fluctuates. It is mainly because when there are few layers, e.g., $ 3 $ layers of the HBDUN, the degree of freedom limits its learning capability. Thus, the sum rate improves with the number of layers. When the number of layers is large, e.g., $ 8 $ layers of the HBDUN, the numerical error caused by the matrix multiplication and inversion becomes large. Then the learning capability of the NN is limited and the sum rate fluctuates. Note that the layers of the CEDUN is the length of the pilots and we can observe that our proposed deep-unfolding NN can save the pilot resources compared to the RLS algorithm. Considering both the sum rate and training time, the optimal choice is $ 5 $ layers for the HBDUN and $ 16 $ layers for the CEDUN. 
	
	\begin{table*}[htbp] \vspace{-0.2em}
		\centering 
		\caption{The CPU running time of the analyzed algorithms.}
		\begin{tabular}{c|cc|ccc}
			\hline
			\multirow{2}*{($ N_{t}^{RF} $,$ K $)} &\multicolumn{2}{c|}{CPU training time (min)}& \multicolumn{3}{c}{CPU testing time (s)} \\
			~ & Joint Design NN & Black-box NN &RLS-SSCA & Joint Design NN & Black-box NN\\
			\hline
			(8,2) &46.82&80.57&2.94&0.26&0.13 \\ \hline
			(16,2) &59.13&96.81&3.11&0.28&0.14 \\ \hline
			(32,2) &76.34&110.45&3.42&0.30&0.14 \\ \hline
			(16,4) &189.23&350.41&11.20&0.96&0.36 \\ \hline
			(32,4) &274.23&435.54&12.81&1.021&0.40 \\ \hline
			(32,6) &394.25&740.61&26.45&1.58&0.76 \\ \hline
		\end{tabular}
		\label{Time} \vspace{-0.2em}
	\end{table*}
	Table \ref{Time} shows the training and testing time of different algorithms with different numbers of RF chains $ N_{t}^{RF} $ and users $ K $. It is observed that the training time of the deep-unfolding NN is much less than the black-box NNs, which indicates that the deep-unfolding NN converges faster. It is because that the CEDUN and HBDUN employ the structure of the RLS and SSCA algorithms, respectively. Moreover, the gap of CPU training time between the deep-unfolding NN and black-box NN increases with $ N_{t}^{RF} $ and $ K $. Furthermore, the testing time of the black-box NN and the joint design NN is much less than that of the RLS-SSCA algorithm, which becomes more obvious as $ N_{t}^{RF} $ and $ K $ increase.
	
	\begin{figure}[t]
		\begin{centering}
			\includegraphics[width=0.43\textwidth]{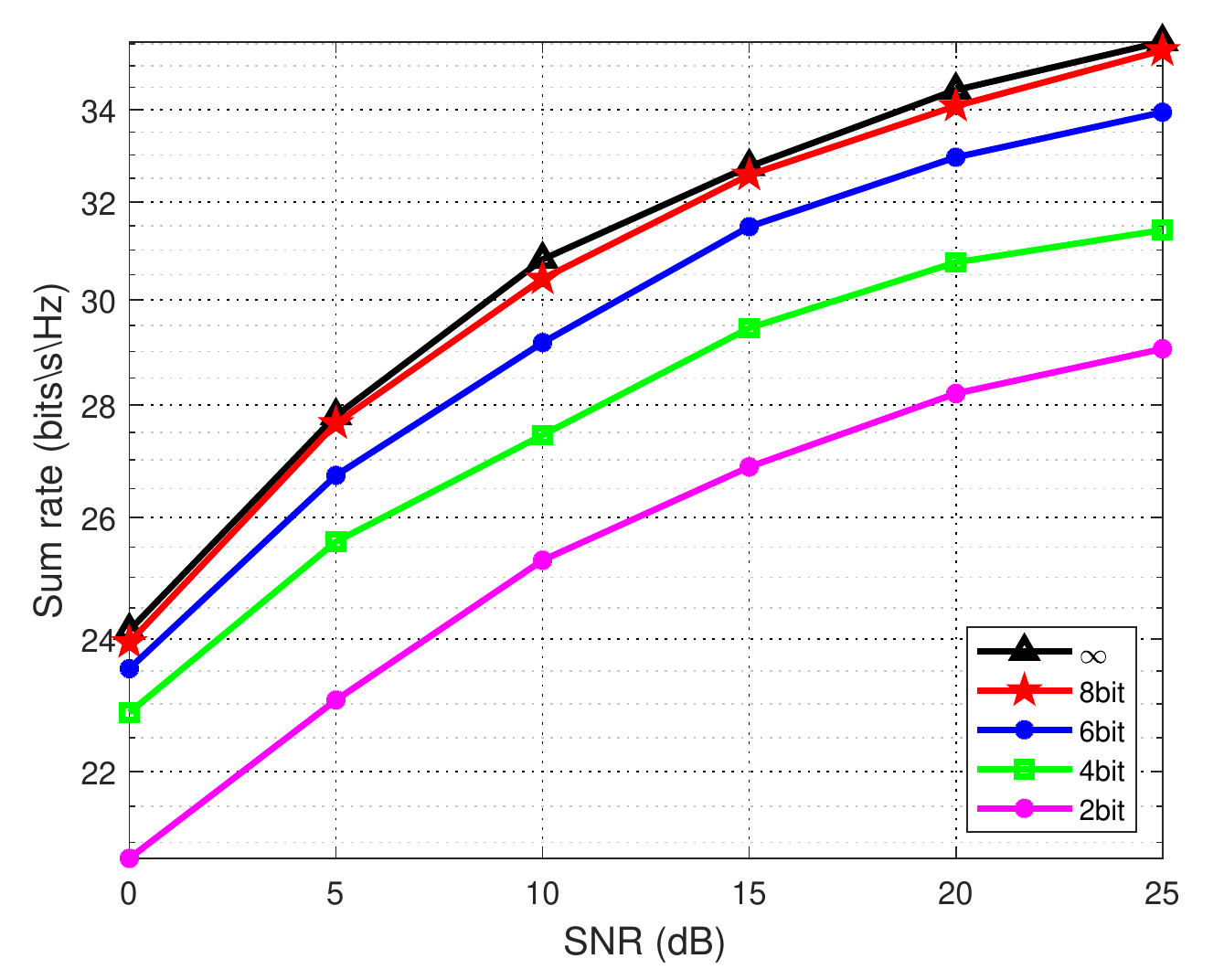}
			\par\end{centering}
		\caption{The sum rate versus the SNR for different numbers of phase shifter quantization
			bits $ Q_{RF} $.}
		\label{analogbit}
	\end{figure}
	{We investigate the impact of finite resolution phase shifters on the deep-unfolding NN. Specifically, we considered a uniform quantization scheme for the analog phase shifters. Fig. \ref{analogbit} presents the sum rate versus the SNR for different numbers of phase shifter
	quantization bits $ Q_{RF} $. It can be seen that the sum rate of the proposed deep-unfolding NN improves with $ Q_{RF} $ as expected. In particular, the performance with $ Q_{RF}  = 8 $ bits can approach the performance with infinite resolution phase shifters.}
	
\begin{figure}[t]
		\begin{centering}
			\includegraphics[width=0.43\textwidth]{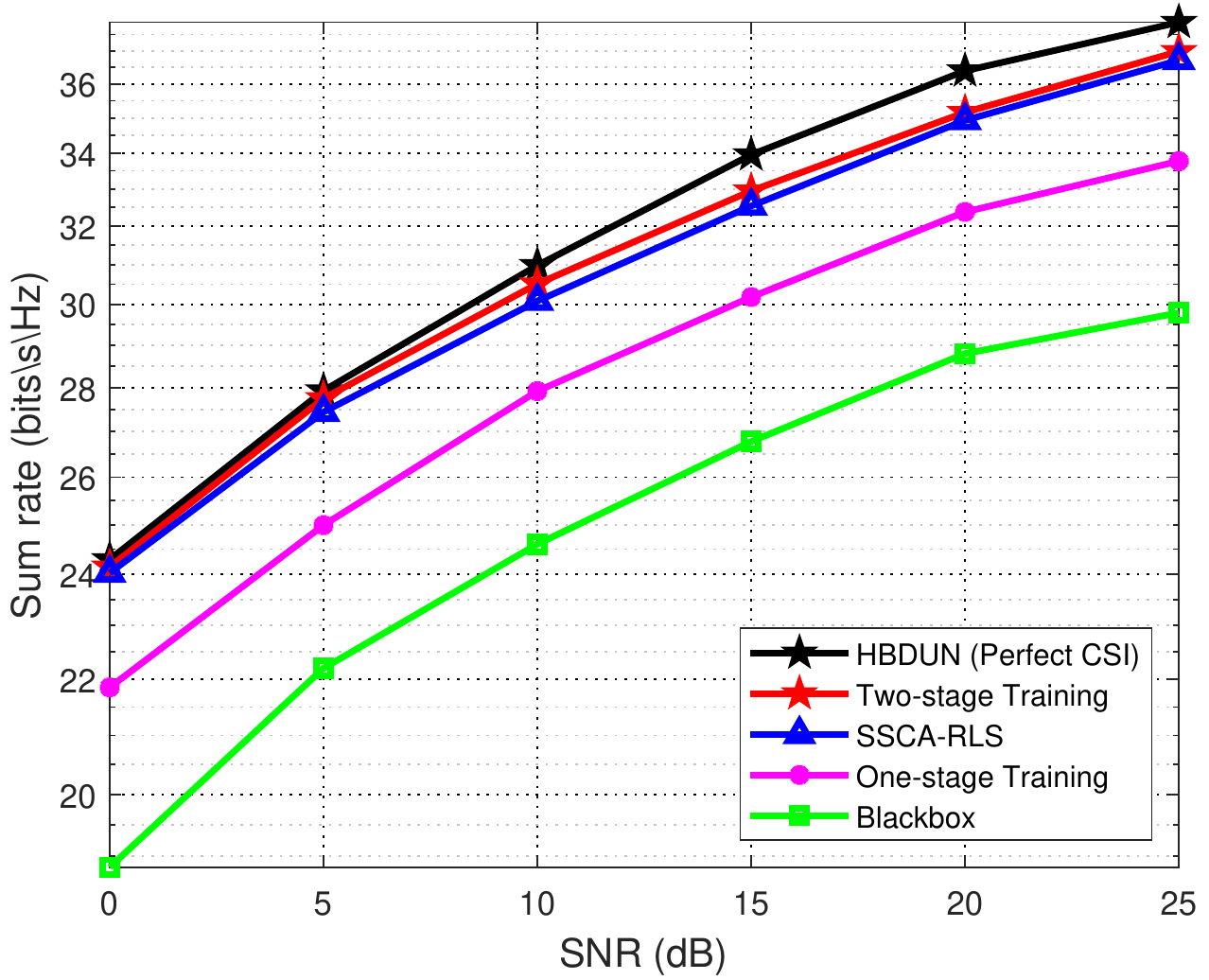}
			\par\end{centering}
		\caption{The sum rate versus the SNR for one/two-stage training and benchmarks.}
		\label{onestage}
	\end{figure}

	{Fig. \ref{onestage}  indicates the sum rate versus the SNR of the one/two-stage training and benchmarks. The curve ``Two-stage Training'' shows the sum rate of the proposed two-stage training while the curve ``One-stage Training'' shows that of one-stage training where we jointly train the whole network shown in Fig.\ref{twotimeframework} (b) in one stage. We see that the performance of two-stage training is better than that of one-stage training as expected, which illustrates the advantage of the proposed two-stage training method.}
    
    {Fig. \ref{generalize}(a) illustrates the generalization ability for different numbers of $ K $ and SNR. We train the joint deep-unfolding NN in the case of $ K=6 $ and $ \textrm{SNR} = 10 $ $ \textrm{dB} $ and test it under different values of $ K $ and SNR, such as $ K=4 $ and $ \textrm{SNR} = 5 $  $\textrm{dB} $. The performance of ``Joint Design NN'' is obtained when the testing scenario is the same as the training scenario while the performance of ``Joint Design NN (mismatch)'' is obtained when there is a mismatch between testing scenario and training scenario. It can be observed that there is only a tiny performance gap between the two curves, which demonstrates that the deep-unfolding NN has satisfactory generalization ability for different numbers of $ K $ and SNR. It is also observed that the performance gap becomes smaller when $ K $ increases.
    
    Fig. \ref{generalize}(b) illustrates the generalization ability for the length of pilot $ L $ and the complex channel gain. We train the joint deep-unfolding NN in the case of $ L $ = 26 and $ \sigma_{\alpha}= 0.1 $ and test it under different values of $ L $ and $ \sigma_{\alpha} $, such as  $ L $ = 22 and $ \sigma_{\alpha}= 0.08 $. Again, there is only a tiny performance gap between the two curves, which demonstrates that the deep-unfolding NN has satisfactory generalization ability for different values of $ L $ and
    $ \sigma_{\alpha} $. The performance gap becomes larger with decreasing $ L $ since the number of layers of the CEDUN decreases with $ L $ and the channel estimation error becomes larger.
	\begin{figure*}[!t]
	\centering
	\subfloat[]{\centering \scalebox{0.46}{\includegraphics{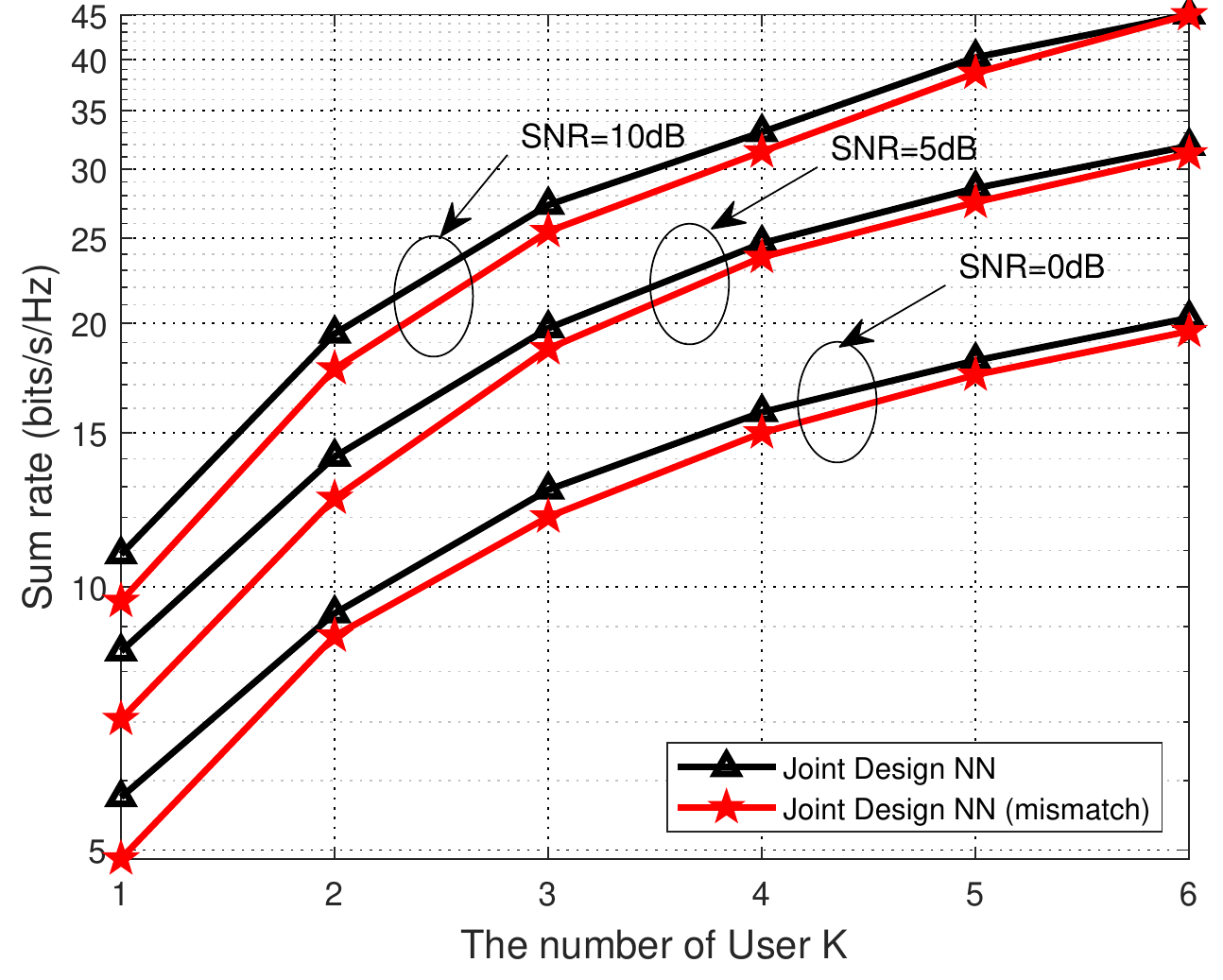}} }
	\subfloat[]{\centering \scalebox{0.46}{\includegraphics{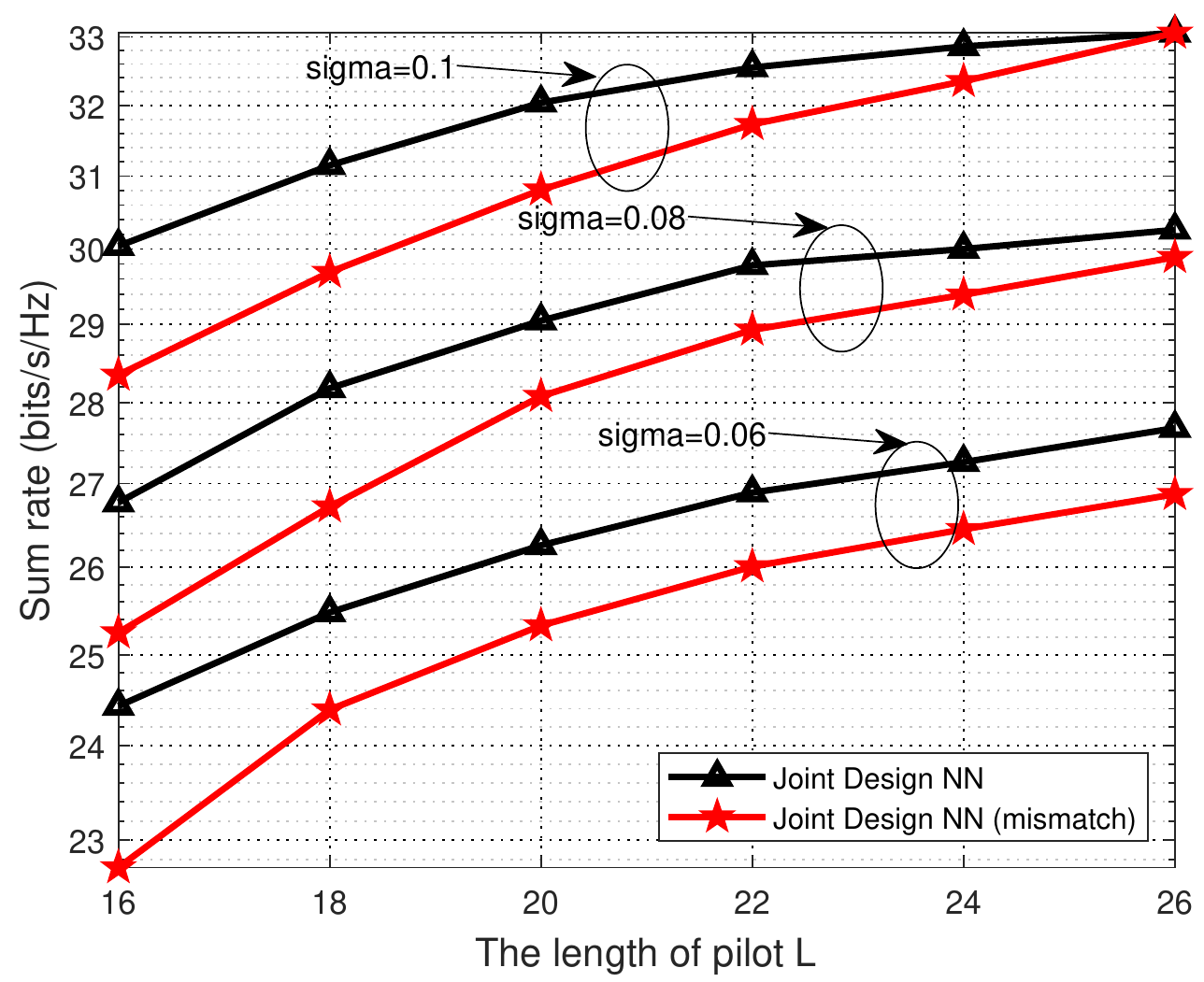}} }
	\caption{The generalization ability:(a) The number of user $ K $ and SNR. (b) The length of pilot $ L $ and the complex gain of rays.}
	\label{generalize}
	\end{figure*}

    \begin{figure}[t]
    	\begin{centering}
    		\includegraphics[width=0.43\textwidth]{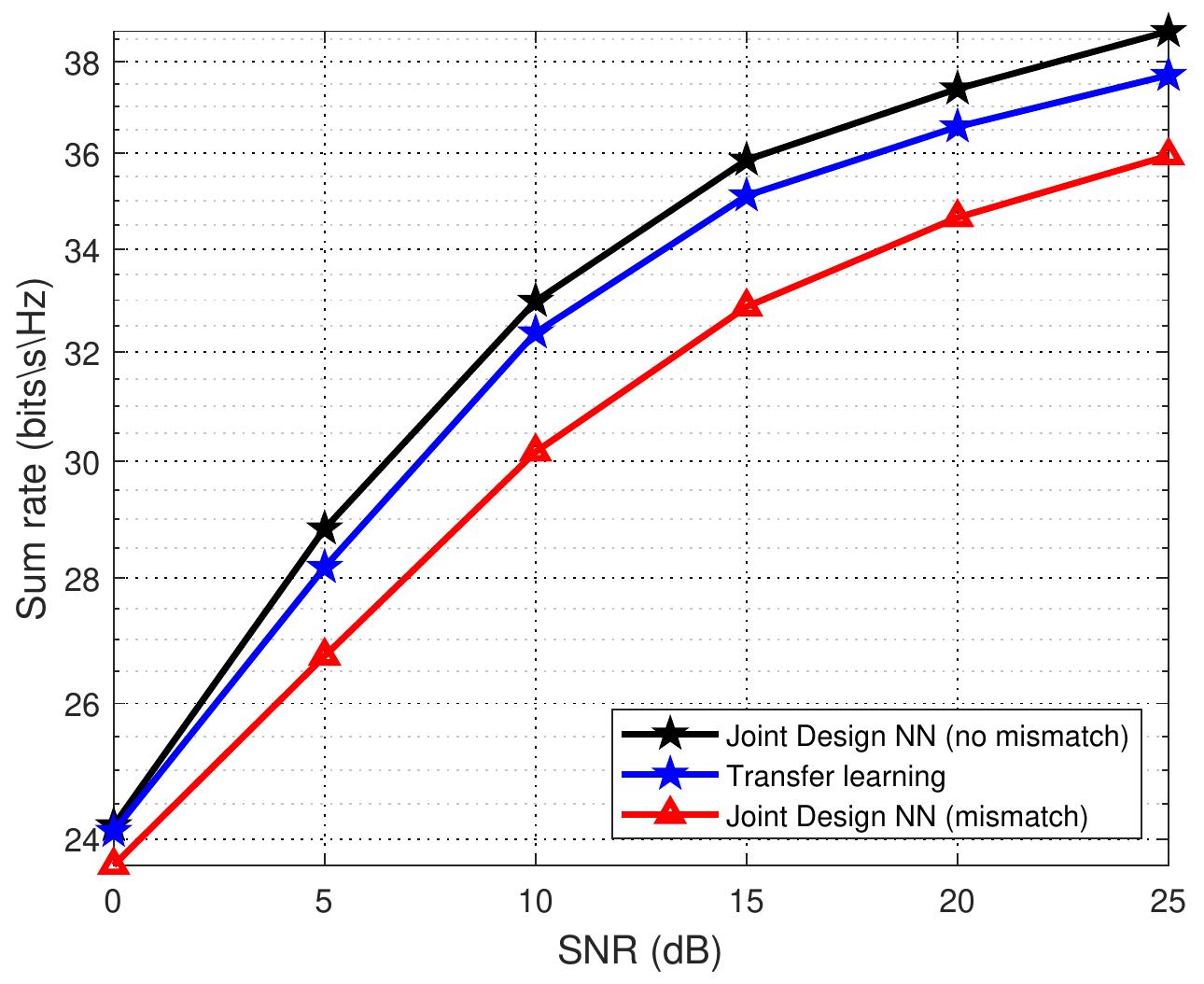}
    		\par\end{centering}
    	\caption{The generalization ability for the channel with different AoAs/AoDs and clusters/rays.}
    	\label{generalize_H}
    \end{figure}
	We show the generalization ability of the deep-unfolding NN for the channel with different AoAs/AoDs and clusters/rays in Fig. \ref{generalize_H}. We train the deep-unfolding NN in the scenario where  $ N_c = 4, N_{ray} = 2,  \phi_{il}^{r} \sim \mathcal{U}(-\pi/3, \pi/3) $ and $ \phi_{il}^{t} \sim \mathcal{U}(-\pi/3, \pi/3) $. The performance of ``Joint Design NN (no mismatch)'' is obtained when the testing scenario is the same as the training scenario. The performance of ``Joint Design NN (mismatch)'' is obtained when $ N_c = 3, N_{ray} = 3,  \phi_{il}^{r} \sim \mathcal{U}(-\pi/2, \pi/2) $ and $ \phi_{il}^{t} \sim \mathcal{U}(-\pi/2, \pi/2) $ in the testing scenario, which is different from that of training scenario. It can be seen that there is a small gap between the performance of ``Joint Design NN (no mismatch)'' and ``Joint Design NN (mismatch)'', which demonstrates that the deep-unfolding NN achieves satisfactory generalization ability for different AoAs/AoDs and numbers of clusters/rays. Furthermore, we employ transfer learning to fine-tune the deep-unfolding NN when the channel statistics change. The ``Transfer learning'' shows the performance of the deep-unfolding NN after transfer learning and is very close to that of ``Joint Design NN (no mismatch)'', which indicates the deep-unfolding NN can adapt to the change of CSI statistics after transfer learning.}
	
    \begin{figure}[t]
 	\begin{centering}
 		\includegraphics[width=0.43\textwidth]{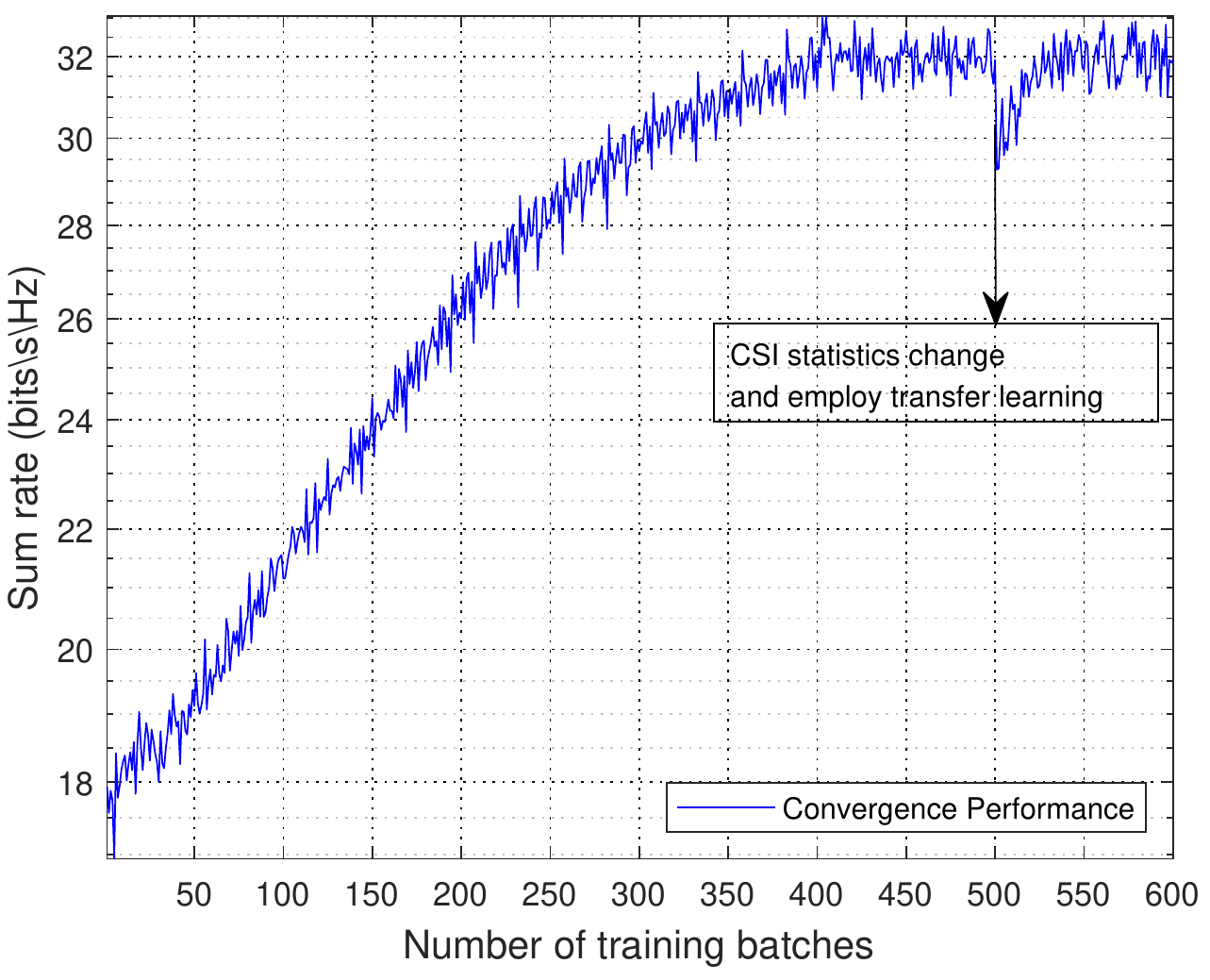}
 		\par\end{centering}
 	\caption{The convergence performance when the CSI statistics change.}
 	\label{transfer}
 	\end{figure}
 	Fig. \ref{transfer} presents the convergence performance of the proposed deep-unfolding NN. It can be seen that when the channel statistics change, the sum rate decreases first, and then increases within several numbers of training batches, which shows that the proposed deep-unfolding NN combined with the method of transfer learning can track the channel variation rapidly. If the channel statistics change fast, we can slightly update the analog beamformers to track the variation.
 	
 	{
	\begin{figure}[t]
 		\begin{centering}
 			\includegraphics[width=0.43\textwidth]{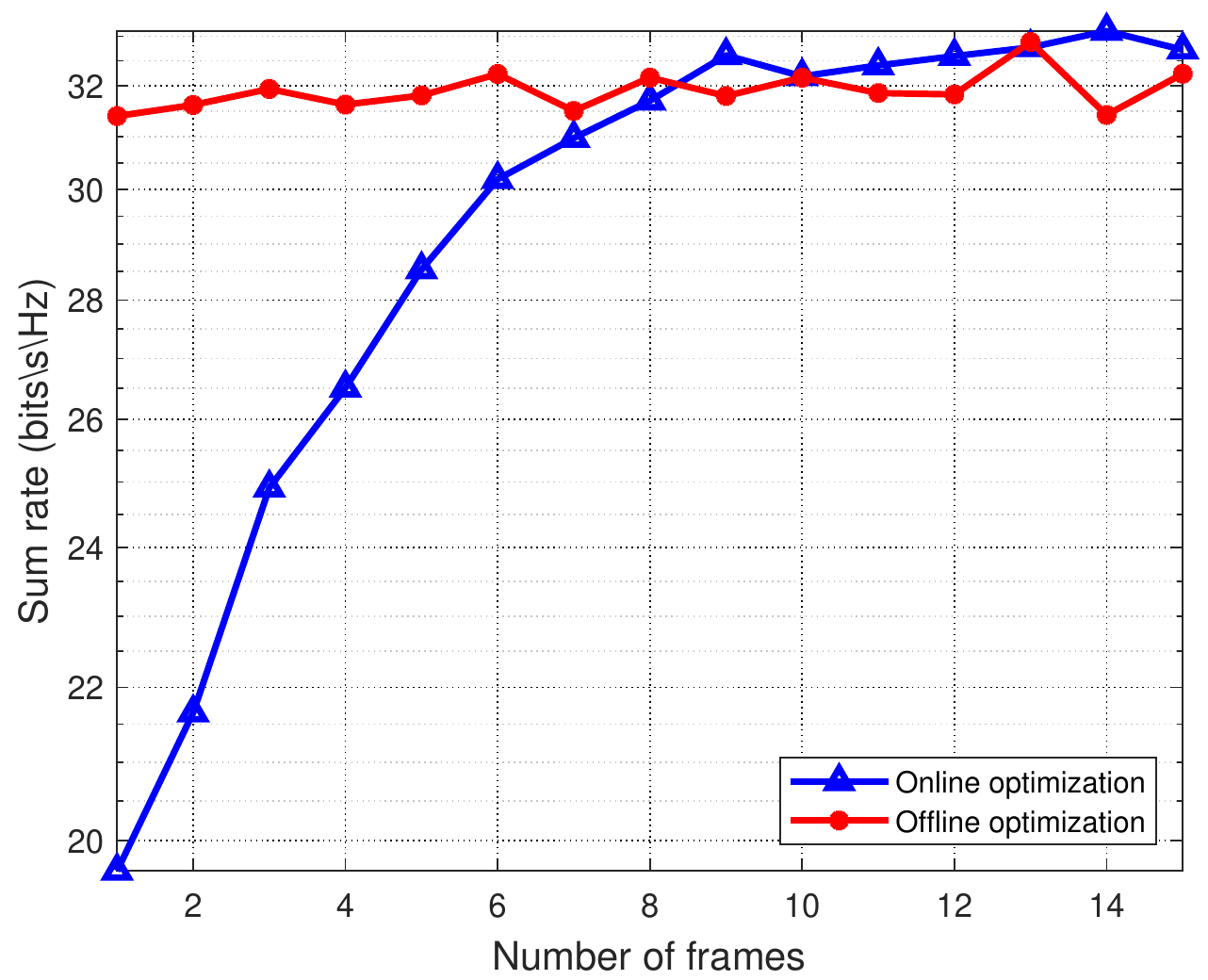}
 			\par\end{centering}
 		\caption{The sum rate of online and offline optimization versus the number of frames.}
 		\label{online_converge}
 	\end{figure}
 	Fig. \ref{online_converge} shows the sum rate of online and offline training versus the number of frames. We collect 50 channel samples each frame to train the deep-unfolding NN in the online training. It can be seen that the sum rate of offline training is stable as the number of frames increases because the deep-unfolding NN is well-trained before data transmission. The sum rate of online optimization gradually improves and eventually converges as the number of frames increases because the deep-unfolding NN is optimized based on the collected channel samples. The converge performance of online training is close to that of offline training. 
 	
	\begin{figure}[t]
 		\begin{centering}
 			\includegraphics[width=0.43\textwidth]{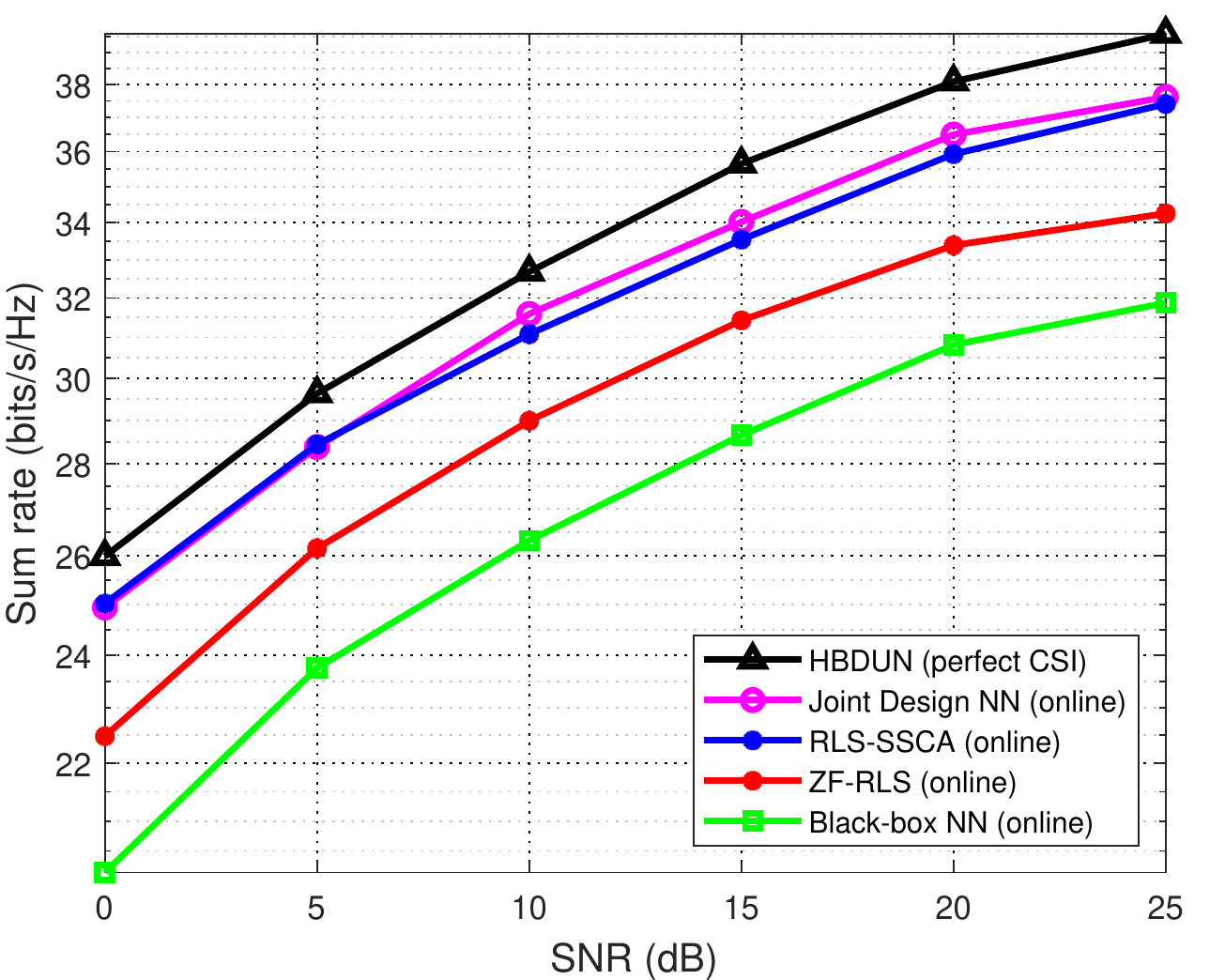}
 			\par\end{centering}
 		\caption{The sum rate of online optimization versus the SNR.}
 		\label{online_SNR}
 	\end{figure}
 
 	Fig. \ref{online_SNR} illustrates the sum rate of the proposed deep-unfolding NN and the benchmarks for different values of SNR. The analog beamformers are optimized online in these algorithms. It is observed that the sum rate increases with SNR for all the algorithms. The performance of the deep-unfolding NN is prominently better than that of the black-box NN. Besides, the proposed joint design deep-unfolding NN achieves comparable performance of the RLS and SSCA algorithm, which indicates the effectiveness of the proposed framework of online optimization.
 	
	\begin{figure}[t]
 		\begin{centering}
 			\includegraphics[width=0.43\textwidth]{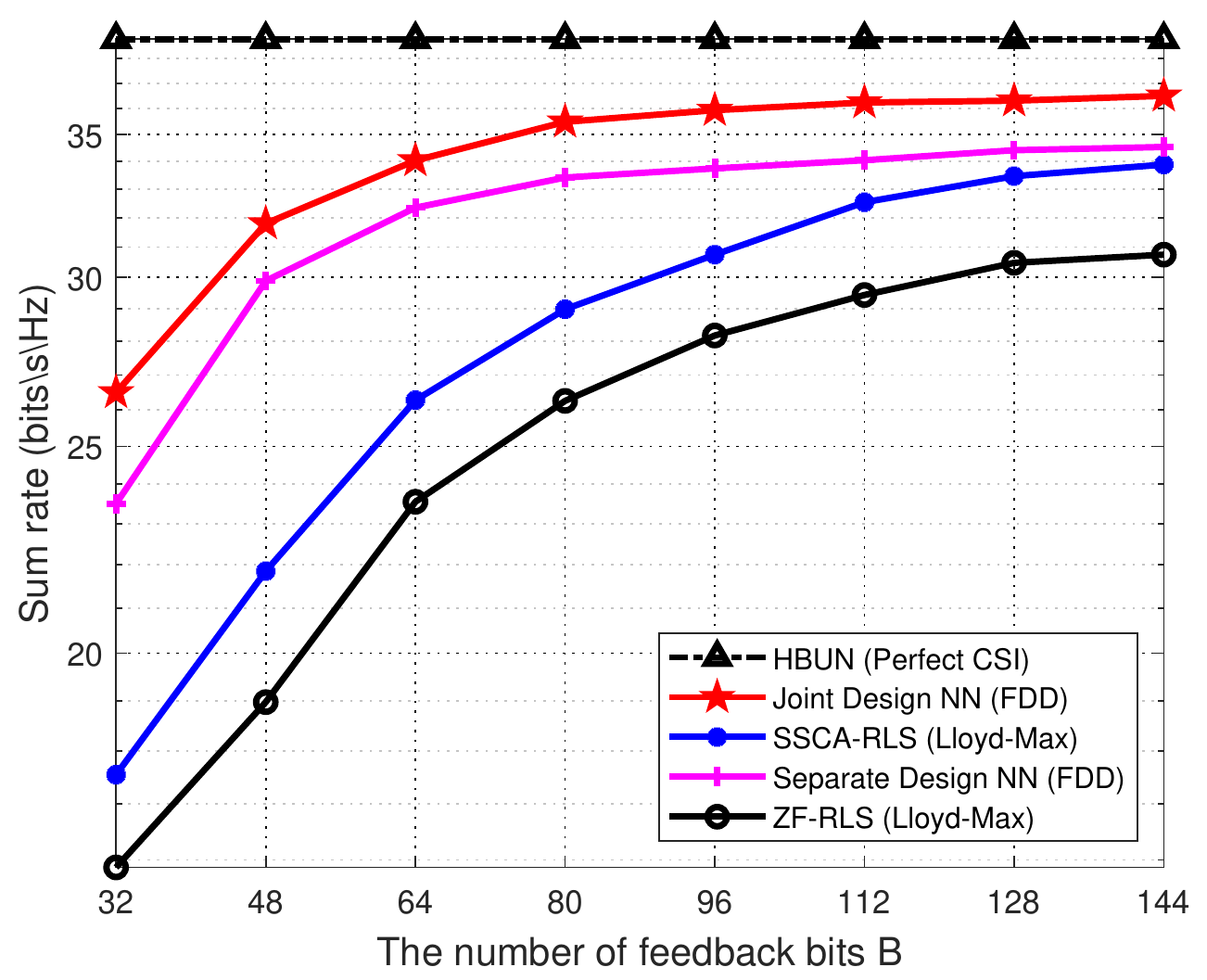}
 			\par\end{centering}
 		\caption{The sum rate versus different numbers of feedback bits.}
 		\label{feedback}
 	\end{figure}
 	Fig. \ref{feedback} illustrates the sum rate versus different numbers of feedback bits in the FDD system. For conventional algorithms, the optimal Lloyd-Max algorithm is employed to quantize the channel parameters $ \{\phi_{il}^{r}, \phi_{il}^{t}, \Re\{\alpha_{il}\}, \Im\{\alpha_{il}\} \}$ and the phase of analog beamformers $ \{\bm{\phi}_{W}, \bm{\phi}_{F}\} $. The performance of ``Separate Design NN (FDD)'' is obtained where the deep-unfolding NNs and the feedback autoencoder are trained separately. The performance of ``Joint Design NN (FDD)'' is obtained where the deep-unfolding NNs and the feedback autoencoder are jointly trained according to the aforementioned training process. It can be observed that the proposed deep-learning NN outperforms the conventional algorithm with the same number of feedback bits. The joint design NN with $ 48 $ feedback bits achieves the same sum rate with the SSCA-RLS algorithm with $ 112 $ feedback bits, which verifies that the proposed deep-learning NN can significantly reduce the number of feedback bits. The joint design NN achieves better performance than the separate design NN, which indicates the effectiveness of our end-to-end joint design deep-learning framework.
 	}
	\section{Conclusion}
	\label{Section7:conclusion}
	In this work, a mixed-timescale deep-unfolding based JCEHB framework has been proposed for hybrid massive MIMO systems. We developed a RLS algorithm induced deep-unfolding NN and an SSCA algorithm induced deep-unfolding NN for channel estimation and hybrid beamforming, respectively. Specifically, we introduced some trainable parameters and non-linear operations to replace the high complexity operations and increase the convergence speed.  { In addition, we propose a mixed-timescale deep-unfolding NN where analog beamformers are optimized online, and extend the framework to the FDD systems where channel feedback is considered.} Furthermore, we analyzed the computational complexity and performance of the proposed deep-unfolding algorithm. The simulation results showed that the proposed deep-unfolding algorithm can outperform the conventional iterative algorithms. For the future study, it is worth generating our deep-unfolding framework to solve more complex wireless systems around the research hot spots, such as multi-cell MIMO, drones and intelligent reflecting surface systems. 
	
	\begin{appendices}
		\label{appendixA}

	\end{appendices}

\bibliographystyle{IEEEtran}
\bibliography{deepunfolding}

\end{document}